\documentclass[aps,pra,twocolumn]{revtex4-2}

\usepackage{graphicx}
\usepackage{dcolumn}
\usepackage{bm}
\usepackage{physics}
\usepackage{float}
\usepackage{qcircuit}
\usepackage{amssymb}
\usepackage{hyperref}
\usepackage{subcaption}
\usepackage{overpic}
\usepackage{multirow}

\hypersetup{
    colorlinks,
    citecolor=blue,
    linkcolor=blue,
    urlcolor=blue
}

\usepackage[whole]{bxcjkjatype}
\usepackage[normalem]{ulem}
\usepackage[dvipsnames]{xcolor}
\usepackage{soul}
\usepackage{pgfplots}
\pgfplotsset{compat=1.18}

\begin{document}

\title{Direct Gradient Computation for Barren Plateaus in Parameterized Quantum Circuits}

\author{Yuhan Yao}
\email{yao@biom.t.u-tokyo.ac.jp}
\author{Yoshihiko Hasegawa}\email{hasegawa@biom.t.u-tokyo.ac.jp}
\affiliation{Department of Information and Communication Engineering,
Graduate School of Information Science and Technology,
The University of Tokyo, Tokyo 113-8656, Japan
}\date{\today}

\begin{abstract}
The barren plateau phenomenon, where the gradients of parametrized quantum circuits become vanishingly small, poses a significant challenge in quantum machine learning.
While previous studies attempted to explain the barren plateau phenomenon using the Weingarten formula, reliance on this formula leads to inaccurate conclusions.
In this study, we consider a unitary operator \(U\) consisting of rotation gates and perform an exact calculation of the expectation required for the gradient computation.
Our approach allows us to obtain the gradient expectation and variance directly.
Our analysis reveals that gradient expectations are not zero, as opposed to the results derived using the Weingarten formula, but depend on the number of qubits in the system.
Furthermore, we demonstrate how the number of effective parameters, circuit depth, and gradient variance are interconnected in deep parameterized quantum circuits.
Numerical simulations further confirm the validity of our theoretical results.
Our approach provides a more accurate framework for analyzing quantum circuit optimization.

\end{abstract}

\maketitle

\textit{Introduction.---}
As quantum machine learning techniques advance, parameterized quantum circuits (PQCs) are increasingly used for various quantum computing tasks.
However, during the training of PQCs, factors such as structure size, circuit depth, and other constraints often cause the optimization process to produce extremely close to zero gradients, effectively stalling the training and degrading overall performance. 
This phenomenon, known as the \textit{barren plateau}, has emerged as a major challenge to the efficient training of PQCs.

For related research on barren plateaus, several studies \cite{Cerezo2021, Anshu2023intro, Leone2024, Uvarov2021gradient, Letcher2024gradient, napp2022gradient, McClean2018, Ragone2024lie, holmes_bp, Larocca_bp, sack_intro, du2022intro, shen2020intro, patti2021intro, wurtz2021intro, larocca2022intro, Russell2017intro, wiersema2020intro, Sciorilli2025intro, Wild2023intro, Palma2023intro, Caro2023intro, Jerbi2023intro, Wiersema2024intro} have employed theoretical analyses based on the principles of gradient expectation and variance. 
The Weingarten formula \cite{weingarten1978, Collins_2022, Grant_bp, yao2025} and its derived methods \cite{Ragone2024lie} were employed to demonstrate that the gradient expectation is always zero under a Haar random distribution \cite{McClean2018}.
The gradient variance characteristics were also investigated for different numbers of qubits and circuit depths.
On the other hand, previous studies \cite{napp2022gradient, Uvarov2021gradient, Letcher2024gradient} effectively estimated the gradient variance by exploring its upper and lower bounds.
In shallow PQCs, studies \cite{Cerezo2021, Anshu2023intro} have shown that using local observables leads to lower variance and more stable convergence than global observables.
Other studies \cite{cerezo2024structure, Leone2024structure} further elaborated that both the circuit structure and the observable affect how parameters contribute to variance and its overall magnitude.
Specifically, different circuit architectures and observables can vary how parameter changes propagate through the circuit, affecting the behavior of the gradient variance.

However, there is a pitfall in the Weingarten formula.
Let us think about a unitary matrix \(U=\sum_i U_i(\boldsymbol{\theta})\) composed of unitary matrices \(U_i\), with \(\theta_i\) being an arbitrary real number, forming the sequence \( \boldsymbol{\theta} \).
As the number of  \(U_i\)  increases, \(U\) gradually approaches the distribution of the Haar measure in the Weingarten formula, and its overall expected value can be expressed as
\begin{align}
\label{eq: weingarten}
     \int d\mu \cdot U^\dagger A U=\mathrm{E}\left[ \left(\sum_i U_i(\boldsymbol{\theta})\right)^\dagger A \left(\sum_i U_i(\boldsymbol{\theta})\right) \right],
\end{align}
where \(d\mu\) represents the Haar probability measure, and \(\int d\mu=1\).
However, in practical scenarios, the expectation should be calculated using the following formula:
\begin{align}
\label{eq: pitfall}
    \mathrm{E}\left[ U(\boldsymbol{\theta})^\dagger A U(\boldsymbol{\theta}) \right] = \sum_{i}\mathrm{E}\left[ U_i(\theta_i)^\dagger\cdot A \cdot U_i(\theta_i) \right].
\end{align}
The difference between the two is that when \( i \neq j \), Eq.~\eqref{eq: weingarten} generates additional cross terms of the form \(\sum_{i,j}\mathrm{E}\left[ U_i(\theta_i)^\dagger\cdot A \cdot U_j(\theta_j) \right]\).
Although the Weingarten formula is a powerful mathematical tool in complex circuits, it can introduce errors. 

In this Letter, we present an exact calculation of the expectation involved in gradient computations.
Our approach systematically calculates both the gradient expectation and variance, providing quantitative insights into the barren plateau phenomenon.
We demonstrate that the gradient expectation is inherently nonzero, challenging previous conclusions derived using the Weingarten formula. Additionally, we establish precise quantitative relationships among gradient variance, circuit depth, and the number of effective parameters. Specifically, we reveal a fundamental scaling law in PQCs: The gradient variance is directly proportional to the ratio of effective parameters. This finding highlights the critical role of parameter efficiency in mitigating barren plateaus. 
\textit{Preliminaries.---}
PQCs are quantum circuits designed with tunable parameters that can be optimized to solve various computational problems, characterized by several key elements: the number of qubits (\(n\)), circuit depth (\(d\)), circuit structure (\(U(\boldsymbol{\theta})\)), initial state (\(\ket{\mathbf{init}}\)), objective function (\(O\)), and optimization process.
The circuit structure can be shown as follows: 
\begin{align}
    U(\boldsymbol{\theta})=\prod_{i=1}^{d}U_{i}(\boldsymbol{\theta}_{i})W_{i},
\end{align}
where \( U_{i}(\boldsymbol{\theta}_{i}) \) represents parameterized quantum gates, such as \( RX \), \( RY \), and \( RZ \), with parameters \( \boldsymbol{\theta}_{i} \) updated during the optimization process. The fixed gates \( W_{i} \), such as \( CX \) or \( CZ \), are used to introduce entanglement within the circuit.

The initial state defines a quantum state $\ket{\mathbf{init}}$, and the circuit ends with measurements to output the expectation values of observables, which are used for optimization.
The loss function, calculated from the measured outputs and serving as the primary metric for optimization, can be expressed as follows:
\begin{align}
\label{eq: loss_function}
\mathcal{L}(\boldsymbol{\theta}) &= \bra{\mathbf{init}}U(\boldsymbol{\theta})^\dagger H U(\boldsymbol{\theta})\ket{\mathbf{init}}\\&=\text{Tr}\{\rho U(\boldsymbol{\theta})^\dagger O U(\boldsymbol{\theta}) \}\notag,
\end{align}
where \(\rho\) denotes \(\ket{\mathbf{init}}\bra{\mathbf{init}}\).
Gradient descent is the core of optimization. It iteratively updates parameters based on gradient calculations to progressively approach the optimal solution for specific application goals.

\begin{figure}
    \centering
    \[
\Qcircuit @C=1em @R=.7em {
  \lstick{\ket{\psi_1}} & \gate{U(\theta_{1})} & \ctrl{1} & \qw & \gate{U(\theta_{5})} & \ctrl{1} & \qw & \gate{U(\theta_{9})} & \ctrl{1} & \meter\\
  \lstick{\ket{\psi_2}} & \gate{U(\theta_{2})} & \ctrl{0} & \ctrl{1} &\gate{U(\theta_{6})} & \ctrl{0} & \ctrl{1} & \gate{{\colorbox{gray}{$U(\theta_{10})$}}} & \ctrl{0} & \ctrl{1} \\
  \lstick{\ket{\psi_3}} & \gate{U(\theta_{3})} & \ctrl{1} & \ctrl{0} & \gate{{\colorbox{gray}{$U(\theta_{7})$}}} & \ctrl{1} & \ctrl{0} & \gate{{\colorbox{gray}{$U(\theta_{11})$}}} & \ctrl{1} & \ctrl{0} \\
  \lstick{\ket{\psi_4}} & \gate{{\colorbox{gray}{$U(\theta_{4})$}}} & \ctrl{0} & \qw & \gate{{\colorbox{gray}{$U(\theta_{8})$}}} & \ctrl{0} & \qw & \gate{{\colorbox{gray}{$U(\theta_{12})$}}} & \ctrl{0} & \qw }
    \]
    \caption{\raggedright Example of the effective parameter. In this circuit, the gray gates do not affect the output, demonstrating that not all trainable parameters are effective for a given task. Although each parameter is trainable, its impact on the output depends on the structure of the circuit and the selected observable values. Some parameters may become redundant or ineffective in influencing the result due to their placement or interaction with other elements in the circuit.}
    \label{fig: effective_parameters_example}
\end{figure}
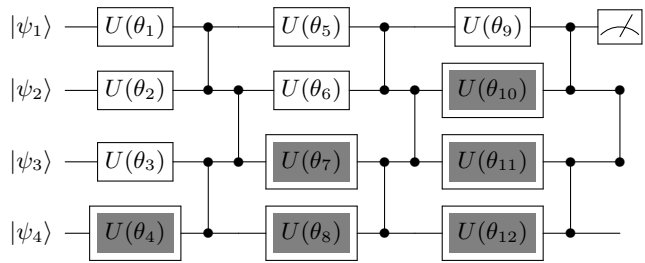

Previous studies \cite{McClean2018, Cerezo2021, Leone2024} have indicated that the barren plateau phenomenon is primarily driven by factors such as the number of qubits, circuit depth, and the choice of observable, based on analyses of the expectation value and variance of the loss function's gradient.
For example, Fig.~\ref{fig: effective_parameters_example} provides an example of how the effect of a parameter on the output depends significantly on the circuit structure and the choice of observable \cite{Leone2024}.
In shallow PQCs, global observables result in exponentially vanishing gradients. However, local observables can effectively mitigate this issue.
Additionally, the number of qubits also significantly influences the gradients in shallow circuits \cite{Cerezo2021}.
In deep PQCs, the Weingarten formula \cite{weingarten1978, Collins_2022} can derive the expectation and variance of the loss function's gradient.
By calculating the gradient of the \(k\)-th parameter, denoted as  $\partial_k \mathcal{L}$, where $\mathcal{L}$ represents the loss function as defined in Eq.\eqref{eq: loss_function}, the gradient expectation can be estimated using Eq.\eqref{eq: weingarten} as follows:
\begin{align}
\mathrm{E}\left[ \partial_k \mathcal{L} \right] =\int d\mu \cdot \partial_k \mathcal{L} =0.
\end{align}
Then, using \(\mathrm{Var}[\partial_{k}\mathcal{L}]=\mathrm{E}[(\partial_{k}\mathcal{L})^2]-\mathrm{E}[\partial_{k}\mathcal{L}]^2\), Ref.~\cite{McClean2018} showed that the gradient variance is given by
\begin{align}
\label{eq: old_var}
\mathrm{Var}[\partial_{k}\mathcal{L}]=\int d\mu \cdot (\partial_k \mathcal{L})^2 \approx\frac{\Tr{O^{2}}\Tr{\rho^{2}}}{2^{3n}-2^n}.
\end{align}
This formula successfully explains the exponential decay relationship between gradient variance and the number of qubits.
However, due to the differences between Eq.\eqref{eq: weingarten}  and Eq.\eqref{eq: pitfall}, this result is actually problematic.
However, it cannot explain the influence of circuit depth and the number of effective parameters, indicating the need for other mathematical tools to explore these relationships further.

\textit{Results.---}The Weingarten formula is limited and fails to capture the relationship between gradient variance, circuit depth, and the number of effective quantum parameters.
Our approach fills this gap by providing a more comprehensive and accurate theoretical framework.
Due to the complexity of the calculations in the Results section, the details are provided in the supplementary material.
To determine the value of \( \mathrm{E}[\partial_k \mathcal{L}]\), we need to compute \(\mathrm{E}\left[U^\dagger AU\right]\).
We cannot simply treat \(U\) as an arbitrary unitary matrix to avoid employing the Weingarten formula.
In an $n$-qubit, 1-depth quantum circuit, we assume a unitary \(U\) is present on the $j$-th qubit.
Here, we utilize the result from Eq.~\eqref{eq: pitfall}, in which \(U\) is constructed from \(RX\), \(RY\), and \(RZ\) gates, so \( U(\theta) \in \left\{RX(\theta), RY(\theta), RZ(\theta)\right\}\otimes I_{\overline{j}}^{\otimes(n-1)}\), where \(\overline{j}\) means except for position $j$ and \(I_{\overline{j}}^{\otimes(n-1)}\) represents an $(n-1)$-fold tensor product of identity matrices, excluding the identity matrix at position $j$.
An explicit expression can be derived as follows:
\begin{align}
    \label{eq: one_exp}
    \mathrm{E}\left[ U(\theta)^\dagger A U(\theta) \right]=\frac{1}{3}\left(A+I_j\otimes\text{Tr}_j A \right),
\end{align}
where \(\text{Tr}_j\) is the partial trace. 

Based on Eq.~\eqref{eq: one_exp}, we can calculate that in the 1-depth case, each qubit is fulfilled with the unitary operator \(U(\theta)\).
At this point, \(U(\boldsymbol{\theta})\) can be expressed as \(\bigotimes_{j=1}^{n} \left\{RX(\theta_{j}), RY(\theta_{j}) ,RZ(\theta_{j})\right\}\), which can be further decomposed as \(\prod_{j=1}^{n}\left\{RX(\theta_{j}), RY(\theta_{j}), RZ(\theta_{j})\right\}\otimes I_{\overline{j}}^{n-1}\).
Thus, the 1-depth result can be obtained by applying this decomposition.
By extending the depth further, we obtain that when \(U(\boldsymbol{\theta})=U(\theta_1, \theta_2,\cdots,\theta_{nd})\), the general case is
\begin{align}
    \label{eq: result_1}
    \mathrm{E}\left[ U^\dagger A  U\right] = \frac{1}{3^n}\sum_{\sigma \in \mathcal{P}(n)}\left\{\left(\frac{4^{|\sigma|}}{3^n}\right)^{d-1}\cdot I_\sigma^{\otimes|\sigma|}\otimes\text{Tr}_\sigma \left\{A\right\}\right\},
\end{align}
where \(\mathcal{P}(n)\) is the power set of \(\{1,2,\cdots,n\}\).
Additionally, \(\mathrm{Tr}_\sigma\) denotes the partial trace over the subsystems indexed by \(\sigma\), and \(|\sigma|\) represents the cardinality of the subset \(\sigma\).
Based on this expression, we can further use it to calculate the expectation value \( \mathrm{E}[\partial_k \mathcal{L}] \).
The \( \mathrm{E}[\partial_k \mathcal{L}] \) in the Weingarten formula is strictly zero. 
However, using Eq.~\eqref{eq: result_1}, we find that the gradient expectation is
\begin{align}
    \label{eq: exp_result}
    \abs{\mathrm{E}[\partial_k \mathcal{L}]} \propto \frac{1}{3^n}.
\end{align}
The expectation calculated in Eq.~\eqref{eq: exp_result} is proportional to $3^{-n}$. While this value becomes extremely small for large $n$, it never reaches zero. This finding contradicts the results from the Weingarten formula, which predicted that the gradient expectation vanishes completely.

To calculate the gradient variance, we first need to evaluate the expectation value $\mathrm{E}\left[ U^\dagger AUB U^\dagger CU\right]$, similar to our earlier analysis of the gradient expectation.
For each operator \(U(\theta) \in \left\{RX(\theta), RY(\theta), RZ(\theta)\right\}\otimes I_{\overline{j}}^{\otimes(n-1)}\) acting within the quantum circuit, the expectation can be derived as follows:
\begin{align}
&\mathrm{E}\left[ U(\theta_{j})^\dagger AU(\theta_{j})BU(\theta_{j})^\dagger C U(\theta_{j}) \right]\\
\leq &\frac{1}{4}\left(ABC+I_j \otimes \text{Tr}_{j}\{AC\}\cdot B\right).\notag
\end{align}

Then, we use the above formula to compute the value at each depth.
Once the value at depth 1 is determined, we incrementally extend the depth to construct the value of the entire quantum circuit.
By progressively increasing the depth, we can observe how the overall outcome evolves with the depth, ultimately leading to the final result:

\begin{widetext}
\begin{align}
    \label{eq: result_2}
    \mathrm{E}\left[ U^\dagger AUB U^\dagger CU\right] = \frac{1}{4^n} \sum_{\sigma \in \mathcal{P}(n)} \left\{ \left(4^{|\sigma|-n}\right)^{d-1} I_\sigma^{\otimes|\sigma|} \otimes \mathrm{Tr}_\sigma \left\{ AC \right\} \cdot B \right\} + O(4^{-n}).
\end{align}
\end{widetext}

The gradient variance can be calculated by the result of \(\mathrm{E}\left[ U^\dagger AUBU^\dagger CU \right]\) in Eq.~\eqref{eq: result_2}. 
This result is difficult to express explicitly.
However, in deep circuits, when $d$ is large enough, we can obtain concise results, as shown below:
\begin{align}
    \label{eq: var_result}
    \mathrm{Var}[\partial_{k}\mathcal{L}] \propto \frac{m}{8^{n}nd},
\end{align}
where \(m\) represents the number of effective parameters explained in Fig.~\ref{fig: effective_parameters_example}.
While the Weingarten formula only reveals the relationship between gradient variance and the number of qubits, our findings further establish that it depends on the number of parameters and the quantum circuit depth, which is shown in the Table.~\ref{tab: Comparison}.

As the depth increases, both $m$ and $d$ grow simultaneously. 
Consequently, the $\frac{m}{d}$ ratio limits the effect of depth on $\mathrm{Var}[\partial_{k}\mathcal{L}]$ in the deep circuit.
As a result, there is virtually no change in the value of $\mathrm{Var}[\partial_{k}\mathcal{L}]$ as the depth increases.

By applying the central limit theorem, it can be demonstrated that the distribution of $\partial_{k}\mathcal{L}$ approaches $\mathcal{N}(\mathrm{E}\left[\partial_{k}\mathcal{L}\right], \mathrm{Var}[\partial_{k}\mathcal{L}])$, where $\mathcal{N}(\mu,\sigma^2)$ is the normal distribution with mean $\mu$ and variance $\sigma^2$.
As $n$ increases, most values approach zero, resulting in $\partial_{k}\mathcal{L}$ tending towards zero. 
This ultimately limits the circuit's performance.

\begin{table}[h]
    \centering
    \begin{tabular}{|c|c|c|}
        \hline
         & previous stduy & our method \\ 
        \hline
        expectation & $\mathrm{E}[\partial_k \mathcal{L}]=0$ & $\abs{\mathrm{E}[\partial_k \mathcal{L}]} \propto \frac{1}{3^n}$ \\
        \hline
        qubits(n) & $\mathrm{Var}[\partial_{k}\mathcal{L}] \propto 8^{-n}$ & $\mathrm{Var}[\partial_{k}\mathcal{L}] \propto 8^{-n}$ \\ 
        \hline
        depths(d) & \textemdash{} & \multirow{4}{*}{$\mathrm{Var}[\partial_{k}\mathcal{L}] \propto \frac{m}{d}$} \\ 
        parameters & \textemdash{} &  \\ 
        input state & $\mathrm{Var}[\partial_{k}\mathcal{L}] \propto \text{Tr}\{\rho^2\}$ & \\
        observables & $\mathrm{Var}[\partial_{k}\mathcal{L}] \propto \text{Tr}\{O^2\}$ & \\ 
        \hline
    \end{tabular}
    \caption{\raggedright Comparison between our method and previous study \cite{McClean2018} in deep PQCs. Quantum circuit depths, parameters, input state, and observables only influence the number of effective parameters.
    }
    \label{tab: Comparison}
\end{table} 
\textit{Numerical Simulations.---}To verify our theoretical calculations, we conduct numerical simulations using the PennyLane toolkit, focusing on standard parameterized quantum circuits as our primary experimental subject. 
In particular, we sample 100 randomly generated quantum circuits and compute the expectations and variances of the overall gradients for each rotation gate's parameters.
This allows us to examine the consistency between theoretical predictions and actual numerical outcomes.
The observable \(O\) used in the measurement process primarily consists of Pauli-Z operators for measured qubits, while the identity operator $I$ is used for those qubits that are not measured.
To gain a deeper understanding of the impact of how the number of qubits ($n$), circuit depth ($d$), the effective number of parameters ($m$), and choice of observable ($O$) influence the gradient distribution, we hold specific variables constant while adjusting others to obtain statistical results and identify patterns across different combinations of variables.
Through this analysis, we aim to confirm our theoretical calculations.

\begin{figure*}[t]
    \centering
    \begin{subfigure}{0.49\textwidth}
        \centering
    \includegraphics[width=\linewidth]{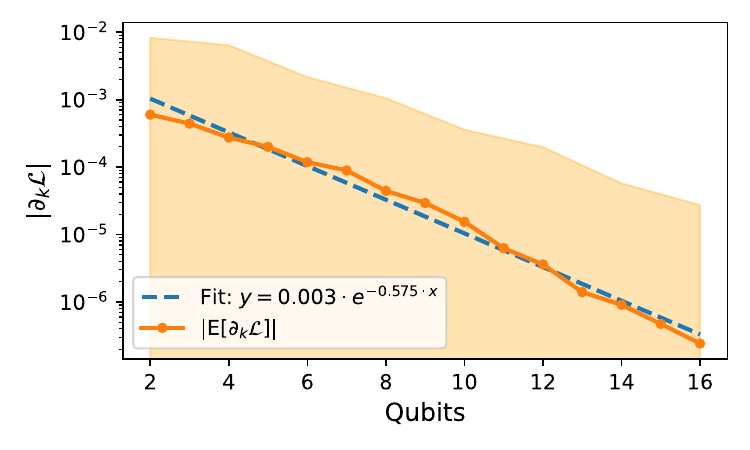}
        \caption{}
        \label{fig: sim_a}
    \end{subfigure}
    \begin{subfigure}{0.49\textwidth}
        \centering
        \includegraphics[width=\linewidth]{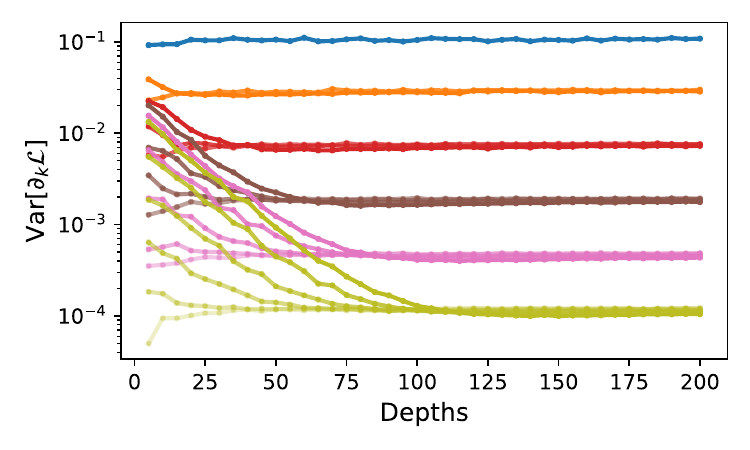}
        \caption{}
        \label{fig: sim_b}
    \end{subfigure}
    \begin{subfigure}{0.49\textwidth}
        \centering
        \includegraphics[width=\linewidth]{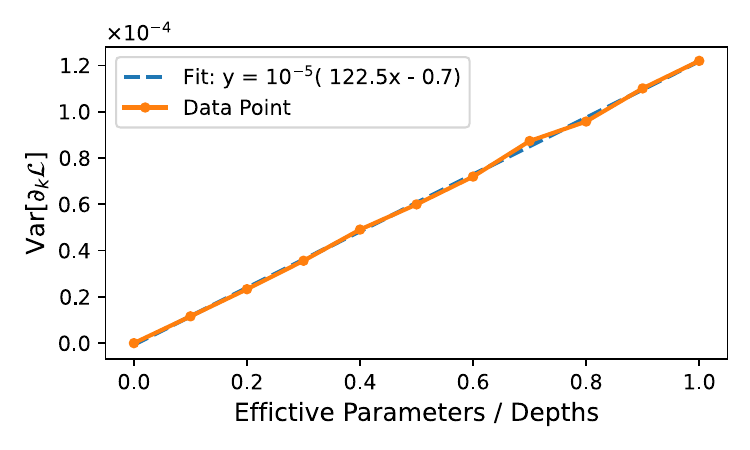}
        \caption{}
        \label{fig: sim_c}
    \end{subfigure}
    \begin{subfigure}{0.49\textwidth}
        \centering
        \includegraphics[width=\linewidth]{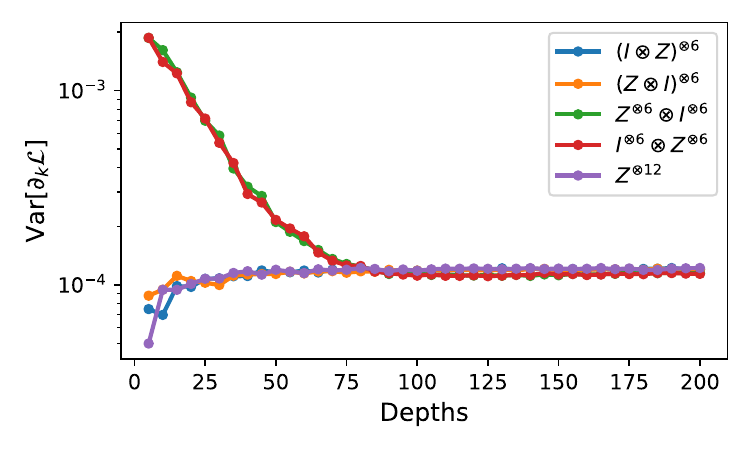}
        \caption{}
        \label{fig: sim_d}
    \end{subfigure}

    \caption{\raggedright The gradient and its variance distributions under different qubit counts, circuit depths, and observables. (a) The gradient distribution for qubit counts ranges from 2 to 16, with circuit depths from 5 to 200, and its expectation is a depth of 200. The light orange region represents the possible values of the gradient. (b) The gradient variance for even qubit counts \{2, 4, 6, 8, 10, 12\} is represented by distinct colors from top to bottom, across observables ranging from $Z^{\otimes 2}$ to $Z^{\otimes n}$, where the color gradually lightens as the number increases, with circuit depths from 5 to 200 in increments of 5. (c) The gradient variance with the effective number of parameters and circuit depth for a 12-qubit system using the \(Z^{\otimes 12}\) observable at a depth of 200. (d) The gradient variance for a 12-qubit system is measured with various observables, including $Z^{\otimes 12}$, \((Z\otimes I)^{\otimes 6}\), \((I\otimes Z)^{\otimes 6}\), \(Z^{\otimes 6}\otimes I^{\otimes 6}\) and \(I^{\otimes 6}\otimes Z^{\otimes 6}\).
    }
    \label{fig:large_image}
\end{figure*}

First, we simulate the gradient expectation to validate Eq.~\eqref{eq: exp_result}.
In this numerical simulation, we employ all qubits from 2 to 16, as shown in Fig.~\ref{fig: sim_a}.
We test the gradients of the loss function in PQCs using the \(Z^{\otimes 2}\) observable, with depths ranging from 5 to 200, measured every 5 depths to obtain the distribution of the entire gradient. 
The expectation value was also measured for the circuit at a depth of 200.
Unlike previous studies \cite{McClean2018, Letcher2024gradient}, we find that while these expectation values are numerically very close to zero, they are not strictly zero, as shown in the expectation row of Table.~\ref{tab: Comparison}.
These results align with our theoretical result in Eq.~\eqref{eq: exp_result}.
Moreover, the results indicate that as the number of qubits increases, the magnitude of these non-zero expectation values decreases. 

In the second simulation, we aim to verify that the gradient variance is independent of the observable but depends on the number of qubits in the deep quantum circuits, following the form of $\mathrm{Var}[\partial_{k}\mathcal{L}] \propto \frac{1}{8^{n}}$ in Eq.\eqref{eq: var_result} rather than $\mathrm{Var}[\partial_{k}\mathcal{L}] \propto \frac{\Tr{O^2}}{2^{3n}-2^n}$ in Eq.\eqref{eq: old_var}.
We consider all even numbers of qubits from 2 to 12 and circuit depths ranging from 5 to 200.
The gradient variance is computed every 5 layers for each qubit count, as shown in Fig.~\ref{fig: sim_b}.
Moreover, we employ a variety of observables, ranging from $Z^{\otimes 2}$ to $Z^{\otimes n}$ for each numerical simulation.
The result demonstrates that for a given qubit count, as the circuit depth becomes sufficiently large, the variance converges to nearly the same value, regardless of which observable is used.
Furthermore, across different qubit counts, the final convergent variance values exhibit an exponential dependence on $8^{-n}$ in both Eq.\eqref{eq: old_var} and Eq.\eqref{eq: var_result}, which has occurred in previous studies \cite{McClean2018, Letcher2024gradient, napp2022gradient}.

In the third numerical simulation, we examine how the gradient variance depends on the effective parameter, as described in Eq.\eqref{eq: var_result}. To do this, we fix the number of qubits at 12 and use $Z^{\otimes 12}$ as the observable.
Then, we replace the rotation gates in the quantum circuit with identity gates or Hadamard gates according to a specified proportion, thereby altering the effective number of parameters in the entire quantum circuit.
The circuit depth ranged from 5 to 200, measured in increments of 5 layers. 
The result in Fig.~\ref{fig: sim_c} demonstrates that the gradient variance stabilizes once the circuit depth becomes sufficiently large and no longer changes significantly.
Its variance value is proportional to the ratio between the effective number of parameters and the circuit depth, corresponding to \(\mathrm{Var}[\partial_{k}\mathcal{L}] \propto \frac{m}{d},\) in Eq.~\eqref{eq: var_result} and Table.~\ref{tab: Comparison}.

In the last numerical simulation, we maintain the 12-qubit setup and analyze how the gradient variance changes with circuit depth under different observables, which is shown in Fig.~\ref{fig: sim_d}.
According to previous theoretical considerations \cite{McClean2018, Ragone2024lie}, due to the gradient variance related to $\mathrm{Tr}(O)$, the observable setting \((Z\otimes I)^{\otimes 6}\) should behave similarly to \(Z^{\otimes 6}\otimes I^{\otimes 6}\).
However, the results shown in Fig.~\ref{fig: sim_d} demonstrate that \((Z\otimes I)^{\otimes 6}\) exhibits a behavior similar to \(Z^{\otimes 12}\) diverging considerably from \(Z^{\otimes 6}\otimes I^{\otimes 6}\).
Our analysis in Eq.~\eqref{eq: var_result} indicates that this discrepancy can be attributed to the ratio of the effective parameters' number to the circuit depth, which \((Z\otimes I)^{\otimes 6}\) is similar to \(Z^{\otimes 12}\).
In contrast, for \(Z^{\otimes 6}\otimes I^{\otimes 6}\), the effective parameter is significantly lower (approximately $nd-21$), resulting in a distinct difference.
This phenomenon suggests that a global observable may be a more effective approach for deep PQCs than a local observable.
 
\textit{Conclusion.---}In this Letter, we have adopted a more straightforward computational strategy that allows us to accurately determine the exact value of \(\mathrm{E}\left[ U^\dagger AU\right]\) and approximate \(\mathrm{E}\left[ U^\dagger AUB U^\dagger CU\right]\).
By doing so, we can effectively analyze the gradient distribution without relying on the Weingarten formula.
This direct method significantly streamlines the analytical process, providing more precise insights into the underlying properties of the gradient behavior and offering a more intuitive framework for subsequent optimization analysis.

Based on our results, we have explained why the gradient expectation is not strictly zero and have identified the relationship between this phenomenon and the number of qubits.
Our findings regarding gradient variance indicate that it depends exponentially on the number of qubits and correlates with the ratio of effective parameters to circuit depth. 
The results are clearly evident in the simulations.
Furthermore, we have provided insights into why the variance for local observables tends to be slightly smaller than that for global observables in deep parameterized quantum circuits.

Our results allow us to avoid potential issues arising from employing additional complex concepts and unclear definitions of the unitary operations, which could otherwise lead to erroneous conclusions.
Our approach is limited to parameterized quantum circuits, where the unitary operators are uniformly chosen from rotation gates. Despite this restriction, it aligns with most numerical simulations currently being performed.
In the future, we intend to examine a broader array of unitary operators and explore the applicability of our method to more complex quantum circuit architectures. 
 
\nocite{*}

\begin{acknowledgments}

This work was supported by JSPS KAKENHI Grant Number JP23K24915.

\end{acknowledgments}

\newpage

\end{document}

\title{Supplementary Material for ``Direct Gradient Computation for Barren Plateaus in Parameterized Quantum Circuits''}

\author{Yuhan Yao}
 \email{yao@biom.t.u-tokyo.ac.jp}
\author{Yoshihiko Hasegawa}%
 \email{hasegawa@biom.t.u-tokyo.ac.jp}
\affiliation{%
 Department of Information and Communication Engineering,
Graduate School of Information Science and Technology,
The University of Tokyo, Tokyo 113-8656, Japan
}

\maketitle
This supplementary material describes the calculations introduced in the main text. The equations and figure numbers are prefixed with ``S'' [for example, Eq.~(S1) or Fig.~S1]. Numbers without this prefix [for example, Eq.~(1) or Fig.~1] refer to items in the main text.

\section{Definitions}

\subsection{Quantum rotation gates}

In this work, \(RX, RY, RZ\) represent single-qubit rotation gates, which rotate quantum states around different axes on the Bloch sphere. They are defined as \( RX(\theta) \), \( RY(\theta) \), \( RZ(\theta) \) represents a quantum gate that rotates the state around the \( X \)-axis, \( Y \)-axis, \( Z \)-axis by an angle \( \theta \) respectively, with its matrix representation given by:
\begin{align}
   \begin{aligned}
   \label{eq: rp}
       RX(\theta) =e^{-i\frac{\theta}{2}X}= \begin{bmatrix}
       \cos\frac{\theta}{2} & -i\sin\frac{\theta}{2} \\
       -i\sin\frac{\theta}{2} & \cos\frac{\theta}{2}
       \end{bmatrix}, \quad 
       RY(\theta) =e^{-i\frac{\theta}{2}Y}= \begin{bmatrix}
       \cos\frac{\theta}{2} & -\sin\frac{\theta}{2} \\
       \sin\frac{\theta}{2} & \cos\frac{\theta}{2}
       \end{bmatrix}, \quad
       RZ(\theta) =e^{-i\frac{\theta}{2}Z}= \begin{bmatrix}
       e^{-i\frac{\theta}{2}} & 0 \\
       0 & e^{i\frac{\theta}{2}}
       \end{bmatrix}.
   \end{aligned}
\end{align}
Here, \( \theta \) is a real number representing the rotation angle. To make the calculation simpler, we change the $RZ$ gate as represented in the following form:
\begin{align}
       RZ(\theta) = e^{-i\frac{\theta}{2}}\begin{bmatrix}
       1 & 0 \\
       0 & e^{i\theta}
       \end{bmatrix}.
\end{align}
This form will not influence the result of \(\mathrm{E}\left[ U^\dagger AU\right]\)\ or \(\mathrm{E}\left[ U^\dagger AUB U^\dagger CU\right]\).

\subsection{Expectation \(\texorpdfstring{\mathrm{E}[f(\theta)]}{E[f(theta)]}\)}

The function \( \mathrm{E}[f(\theta)] \) is used to describe the expectation of a function \( f(\theta) \) over the parameter \( \theta \), and is defined as:
\begin{align}
\mathrm{E}[f(\theta)] = \int f(\theta) p(\theta) d\theta,
\end{align}
where:
\begin{enumerate}
    \item \( p(\theta) \) is a probability distribution over \( \theta \).
    \item \( f(\theta) \) is a measurable function of \( \theta \).
\end{enumerate}

\( \mathrm{E}[f(\theta)] \) represents the weighted average of \( f(\theta) \) under the probability distribution \( p(\theta) \), commonly used in quantum computation to describe the statistical properties of parameterized gate operations.

Since \( p(\theta) \) is uniformly distributed over \( \mathbb{R} \), to avoid issues \(\mathrm{E}_1\neq\mathrm{E}_2\) as follows:
\begin{align}
    \mathrm{E}_1\left[\Tr{Ry(\theta)}\right]=\frac{1}{2\pi}\int_{0}^{2\pi}2\cos{\frac{\theta}{2}}\,d\theta=0,\\
    \mathrm{E}_2\left[\Tr{Ry(\theta)}\right]=\frac{1}{2\pi}\int_{-\pi}^{\pi}2\cos{\frac{\theta}{2}}\,d\theta=\frac{4}{\pi}.
\end{align}

we introduce a period \( T=4\pi \) to constrain \( \theta \) within the interval \( [-2\pi, 2\pi) \). Under this setup, the expectation \( \mathrm{E}[f(\theta)] \) is defined as:
\begin{align}
\mathrm{E}[f(\theta)] = \frac{1}{4\pi} \int_{-2\pi}^{2\pi} f(\theta) \, d\theta,
\end{align}
where:
\begin{enumerate}
    \item \( T=4\pi \) represents the period, ensuring \( \theta \) is confined within a finite interval, and ensure that \(\mathrm{E}[f(\theta)]\) is a fixed value.
    \item \( \frac{1}{4\pi} \) is the normalization factor to maintain the consistency of the probability distribution.
\end{enumerate}

This definition avoids issues arising from a uniform distribution over the unbounded interval, such as divergence or non-integrability, over the unbounded interval \( \mathbb{R} \).

Based on the above definition, we can obtain the following formulas：
\begin{align}
    &\mathrm{E}\left[\theta\right] = 0, \quad \mathrm{E}\left[\theta^2\right]=\frac{4\pi^2}{3}.\\
    \label{eq: one}
    &\mathrm{E}\left[\cos\frac{\theta}{2}\right]=\mathrm{E}\left[\sin\frac{\theta}{2}\right] = 0.\\
    \label{eq: two}
    &\mathrm{E}\left[\cos^2\frac{\theta}{2}\right]=\mathrm{E}\left[\sin^2\frac{\theta}{2}\right] = \frac{1}{2}, \quad \mathrm{E}\left[\cos\frac{\theta}{2}\sin\frac{\theta}{2}\right] = 0.\\  
    \label{eq: three}
    &\mathrm{E}\left[\cos^3\frac{\theta}{2}\right]= \mathrm{E}\left[\cos^2\frac{\theta}{2}\sin\frac{\theta}{2}\right] = \mathrm{E}\left[\cos\frac{\theta}{2}\sin^2\frac{\theta}{2}\right] =\mathrm{E}\left[\sin^3\frac{\theta}{2}\right] = 0.\\ 
    \label{eq: four}
    &\mathrm{E}\left[\cos^4\frac{\theta}{2}\right] =\mathrm{E}\left[\sin^4\frac{\theta}{2}\right] = \frac{3}{8}, \quad \mathrm{E}\left[\cos^3\frac{\theta}{2}\sin\frac{\theta}{2}\right] = \mathrm{E}\left[\cos\frac{\theta}{2}\sin^3\frac{\theta}{2}\right] = 0, \quad \mathrm{E}\left[\cos^2\frac{\theta}{2}\sin^2\frac{\theta}{2}\right] = \frac{1}{8}.\\ 
    \label{eq: exp}
    &\mathrm{E}\left[e^{ik\theta}\right] =
\begin{cases} 
1, & \text{if } k = 0, \\
0, & \text{if } k \neq 0.
\end{cases}
\end{align}

\subsection{Loss function's gradient}

For parameterized quantum circuits, we define its loss function as:
\begin{align}
\mathcal{L}(\boldsymbol{\theta})= \bra{\mathbf{init}}U(\boldsymbol{\theta})^\dagger H U(\boldsymbol{\theta})\ket{\mathbf{init}}=\text{Tr}\{\rho U(\boldsymbol{\theta})^\dagger O U(\boldsymbol{\theta}) \},
\end{align}
where \(\rho\) presents the \(\ket{\mathbf{init}}\bra{\mathbf{init}}\).

For the \( k \)-th parameter, the gradient of the loss function can be expressed as:
\begin{align}
    \partial_k \mathcal{L} &= \frac{\partial}{\partial \theta_k} \text{Tr} \{\rho U^\dagger O U\}\\ 
    &= \text{Tr} \{O_+ \cdot \partial_k (RP_k(\theta_k)\cdot \rho_- \cdot RP_k(\theta_k)^\dagger)\}\\
    \label{eq: rp_diff}
    &=\text{Tr} \left\{O_+ \cdot \left[\partial_k (RP_k(\theta_k))\cdot \rho_- RP_k(\theta_k)^\dagger + RP_k(\theta_k) \rho_-\cdot \partial_k (RP_k(\theta_k))^\dagger\right]\right\}\\
    \label{eq: rp_diffs}
    &= \frac{i}{2}\text{Tr} \left\{O_+ \cdot \left[RP_k(\theta_k) \rho_- RP_k(\theta_k)^\dagger, P_k\right]\right\},
\end{align}
where:
\begin{enumerate}
    \item \( \theta_k \) is the \( k \)-th parameter.
    \item For \(P\in \{X, Y, Z\}\), \(RP(\theta)\) comes from Eq.~\eqref{eq: rp}.
    \item The partial of \(RP_k(\theta_k)\) is calculated as:
    \begin{align}
        \partial_k (RP_k(\theta_k)) = \partial_k (e^{-i\frac{\theta}{2}P_k})=-\frac{i}{2}P_ke^{-i\frac{\theta}{2}P_k}=-\frac{i}{2}P_k\cdot RP_k(\theta_k).
    \end{align}
    \item \(\rho_- = U_- \rho U_-^\dagger \).
    \item \(O_+ = U_+^\dagger O U_+\).
\end{enumerate}

To illustrate this process more intuitively, we refer to the figure below, where the yellow part represents \( U_- \), the pink part represents \( U_+ \), and \(RP_k\) is white:

\begin{figure}[h]
    \centering
    \[
\Qcircuit @C=1em @R=.7em {
  \lstick{\ket{\psi_1}} & \gate{{\colorbox{yellow}{$U(\theta_{1})$}}} & \ctrl{1} & \qw & \gate{{\colorbox{yellow}{$U(\theta_{5})$}}} & \ctrl{1} & \qw & \gate{{\colorbox{pink}{$U(\theta_{9})$}}} & \ctrl{1} & \qw\\
  \lstick{\ket{\psi_2}} & \gate{{\colorbox{yellow}{$U(\theta_{2})$}}} & \ctrl{0} & \ctrl{1} &\gate{RP_{k}(U_k))} & \ctrl{0} & \ctrl{1} & \gate{{\colorbox{pink}{$U(\theta_{10})$}}} & \ctrl{0} & \ctrl{1} \\
  \lstick{\ket{\psi_3}} & \gate{{\colorbox{yellow}{$U(\theta_{3})$}}} & \ctrl{1} & \ctrl{0} & \gate{{\colorbox{pink}{$U(\theta_{7})$}}} & \ctrl{1} & \ctrl{0} & \gate{{\colorbox{pink}{$U(\theta_{11})$}}} & \ctrl{1} & \ctrl{0} \\
  \lstick{\ket{\psi_4}} & \gate{{\colorbox{yellow}{$U(\theta_{4})$}}} & \ctrl{0} & \qw & \gate{{\colorbox{pink}{$U(\theta_{8})$}}} & \ctrl{0} & \qw & \gate{{\colorbox{pink}{$U(\theta_{12})$}}} & \ctrl{0} & \qw }
    \]
    \caption{An example of $U_+$ and $U_-$.}
    \label{fig: U+-}
\end{figure}
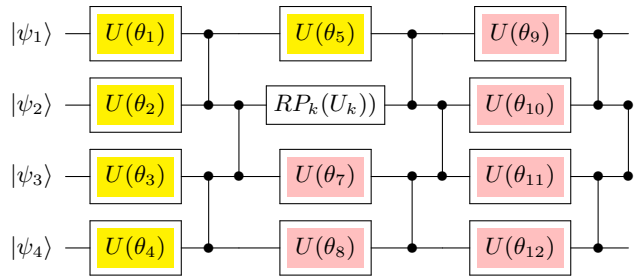

Since we can combine the terms \( RP_k(\theta_k) \) and \( \rho_- \), the final form of \( \partial_k \mathcal{L} \) can be expressed as:
\begin{align}
    \partial_k \mathcal{L} = \frac{i}{2}\text{Tr} \left\{O_+ \cdot \left[\rho_-, P_k\right]\right\}.
\end{align}

\section{Derivation of Eq.~\UAUpoint{}\label{app: 22}}

In our analysis, we start with the most fundamental case of \(\mathrm{E}\left[ U^\dagger A U \right]\), which involves a single-qubit, single-depth, and single-parameter setup.
Next, we extend this to $n$-qubit systems with single-depth and single-parameter configurations.
Subsequently, we generalize to $n$-qubit systems with a single-depth but $n$-parameter case. Finally, we consider the most comprehensive scenario involving $n$-qubit systems with $d$-depth and $nd$-parameter setups.

First, we need to clarify one point: the set of unitary operators is 
\begin{align}
    U(\theta)\in \left\{RX(\theta), RY(\theta), RZ(\theta)\right\},
\end{align}
instead of 
\begin{align}
U'(\theta)=\frac{1}{3}\left( RX(\theta)+RY(\theta)+RZ(\theta) \right).
\end{align}
Although \( U'(\theta) \) is also a unitary operator, it introduces cross-phases when calculating \( \mathrm{E}\left[ U^\dagger A U \right] \), which does not align with the actual physical scenario. Therefore, \( \mathrm{E}\left[ U^\dagger A U \right] \) must be defined using \( U(\theta) \).  

For a single-qubit circuit, there are three possible cases, corresponding to the use of \( RX \), \( RY \), and \( RZ \). Thus, \( \mathrm{E}\left[ U^\dagger A U \right] \) can be expressed as:
\begin{align}
\mathrm{E}\left[ U(\theta)^\dagger A U(\theta) \right]=\frac{1}{3}\big( \mathrm{E}\left[ RX(\theta)^\dagger \cdot A\cdot RX(\theta) \right]
+\mathrm{E}\left[ RY(\theta)^\dagger \cdot A\cdot RY(\theta) \right]
+\mathrm{E}\left[ RZ(\theta)^\dagger \cdot A\cdot RZ(\theta) \right]\big).
\end{align}
Next, we calculate for each case, and we need first to define the form of matrix $A$ as follows:
\begin{align}
    A = \begin{bmatrix}
        a_{11} & a_{12} \\
        a_{21} & a_{22}
    \end{bmatrix}.
\end{align}
\begin{enumerate}
    \item For \( RX \): 
    \begin{align}
        &\mathrm{E}\left[ RX(\theta)^\dagger \cdot A\cdot RX(\theta) \right]\\ =& \mathrm{E}\left[ \begin{bmatrix}
   \cos\frac{\theta}{2} & -i\sin\frac{\theta}{2} \\
   -i\sin\frac{\theta}{2} & \cos\frac{\theta}{2}
   \end{bmatrix}^\dagger \cdot \begin{bmatrix}
   a_{11} & a_{12} \\
   a_{21} & a_{22}
   \end{bmatrix}\cdot \begin{bmatrix}
   \cos\frac{\theta}{2} & -i\sin\frac{\theta}{2} \\
   -i\sin\frac{\theta}{2} & \cos\frac{\theta}{2}
   \end{bmatrix} \right]\\
   \label{eq: rx_original}
   =&\mathrm{E}\begin{bmatrix}
\left(a_{11}\cos^2\frac{\theta}{2} + a_{22}\sin^2\frac{\theta}{2}\right) - i(a_{12} + a_{21})\cos\frac{\theta}{2}\sin\frac{\theta}{2} &
\left(a_{12}\cos^2\frac{\theta}{2} + a_{21}\sin^2\frac{\theta}{2}\right) + i(a_{11} - a_{22})\cos\frac{\theta}{2}\sin\frac{\theta}{2} \\
\left(a_{21}\cos^2\frac{\theta}{2} + a_{12}\sin^2\frac{\theta}{2}\right) + i(a_{11} - a_{22})\cos\frac{\theta}{2}\sin\frac{\theta}{2} &
\left(a_{22}\cos^2\frac{\theta}{2} + a_{11}\sin^2\frac{\theta}{2} \right)+ i(a_{12} + a_{21})\cos\frac{\theta}{2}\sin\frac{\theta}{2}
\end{bmatrix}\\
\label{eq: rx_after}
=& \frac{1}{2}\begin{bmatrix}
a_{11}+ a_{22} &
a_{12}+ a_{21} \\
a_{21}+ a_{12} &
a_{22}+ a_{11}
\end{bmatrix}.
\end{align}

\item For \( RY \): 
    \begin{align}
        &\mathrm{E}\left[ RY(\theta)^\dagger \cdot A\cdot RY(\theta) \right]\\ =& \mathrm{E}\left[ \begin{bmatrix}
       \cos\frac{\theta}{2} & -\sin\frac{\theta}{2} \\
       \sin\frac{\theta}{2} & \cos\frac{\theta}{2}
       \end{bmatrix}^\dagger \cdot \begin{bmatrix}
   a_{11} & a_{12} \\
   a_{21} & a_{22}
   \end{bmatrix}\cdot \begin{bmatrix}
       \cos\frac{\theta}{2} & -\sin\frac{\theta}{2} \\
       \sin\frac{\theta}{2} & \cos\frac{\theta}{2}
       \end{bmatrix} \right]\\
       \label{eq: ry_original}
   =&\mathrm{E}
\begin{bmatrix}
\left(a_{11}\cos^2\frac{\theta}{2} + a_{22}\sin^2\frac{\theta}{2}\right) + (a_{12} + a_{21})\cos\frac{\theta}{2}\sin\frac{\theta}{2} &
\left(a_{12}\cos^2\frac{\theta}{2} - a_{21}\sin^2\frac{\theta}{2}\right) + (a_{22} - a_{11})\cos\frac{\theta}{2}\sin\frac{\theta}{2} \\
\left(a_{21}\cos^2\frac{\theta}{2} - a_{12}\sin^2\frac{\theta}{2}\right) + (a_{22} - a_{11})\cos\frac{\theta}{2}\sin\frac{\theta}{2} &
\left(a_{22}\cos^2\frac{\theta}{2} + a_{11}\sin^2\frac{\theta}{2}\right) - (a_{12} + a_{21})\cos\frac{\theta}{2}\sin\frac{\theta}{2}
\end{bmatrix}\\
\label{eq: ry_after}
=& \frac{1}{2}\begin{bmatrix}
a_{11}+ a_{22} &
a_{12}- a_{21} \\
a_{21}- a_{12} &
a_{22}+ a_{11}
\end{bmatrix}.
    \end{align}

\item For \( RZ \):
\begin{align}
    &\mathrm{E}\left[ RZ(\theta)^\dagger \cdot A\cdot RZ(\theta) \right]\\
    =& \mathrm{E}\left[ \begin{bmatrix}
       1 & 0 \\
       0 & e^{i\theta}
       \end{bmatrix}^\dagger \cdot \begin{bmatrix}
   a_{11} & a_{12} \\
   a_{21} & a_{22}
   \end{bmatrix}\cdot \begin{bmatrix}
       1 & 0 \\
       0 & e^{i\theta}
       \end{bmatrix} \right]\\
       \label{eq: rz_original}
    =& \mathrm{E}\left[ \begin{bmatrix}
   a_{11} & a_{12}e^{i\theta} \\
   a_{21}e^{-i\theta} & a_{22}
   \end{bmatrix} \right]\\
   \label{eq: rz_after}
   =& \begin{bmatrix}
   a_{11} & 0 \\
   0 & a_{22}
   \end{bmatrix}.
\end{align}
\end{enumerate}
From the Eq.~\eqref{eq: rx_original} to Eq.\eqref{eq: rx_after} and Eq.~\eqref{eq: ry_original} to Eq.\eqref{eq: ry_after} use the conclusion of Eq.\eqref{eq: two}. Similarly, Eq.~\eqref{eq: rz_original} to Eq.\eqref{eq: rz_after} uses the result of Eq.\eqref{eq: exp}.

Finally, to avoid directly using the symbol \( a_{ij} \), we can rewrite the result of \( \mathrm{E}\left[ U^\dagger A U \right] \) as:
\begin{align}
\mathrm{E}\left[ U(\theta)^\dagger A U(\theta) \right]=&\frac{1}{3}\big( \mathrm{E}\left[ RX(\theta)^\dagger \cdot A\cdot RX(\theta) \right]
+\mathrm{E}\left[ RY(\theta)^\dagger \cdot A\cdot RY(\theta) \right]
+\mathrm{E}\left[ RZ(\theta)^\dagger \cdot A\cdot RZ(\theta) \right]\big)\\
=&\frac{1}{3}\bigg( \frac{1}{2}\begin{bmatrix}
a_{11}+ a_{22} &
a_{12}+ a_{21} \\
a_{21}+ a_{12} &
a_{22}+ a_{11}
\end{bmatrix} + \frac{1}{2}\begin{bmatrix}
a_{11}+ a_{22} &
a_{12}- a_{21} \\
a_{21}- a_{12} &
a_{22}+ a_{11}
\end{bmatrix} + \begin{bmatrix}
   a_{11} & 0 \\
   0 & a_{22}
   \end{bmatrix} \bigg)\\
   \label{eq: u}
=&\frac{1}{3}\begin{bmatrix}
   a_{11}+a_{22} & a_{12} \\
   a_{21} & a_{22}+a_{11}
   \end{bmatrix} = \frac{1}{3}\left(A+I \cdot \Tr{A}\right).
\end{align}

This result, though simple in form, avoids the explicit use of \( a_{ij} \) and instead represents \( \mathrm{E}\left[ U(\theta)^\dagger A U(\theta) \right] \) entirely in terms of \( A \).

It is very challenging to directly compute the \( \mathrm{E}_{ij}\left[ U(\theta_{ij})^\dagger A U(\theta_{ij}) \right] \) for \( n \)-parameter systems at single-depth based solely on the previous results.
However, for the single-parameter case, the single-qubit scenario can be regarded as a special case of \( n \)-qubit systems.
Therefore, we can calculate intermediate states step by step.
Specifically, we start with the \( 2 \)-qubit scenario, where \( U_{ij} \) can be expressed as:
\begin{align}
    U_{ij}(\theta_{ij}) \in \left\{RX_j(\theta_{ij})\otimes I_{\overline{j}}^{n-1}, RY_j(\theta_{ij})\otimes I_{\overline{j}}^{n-1}, RZ_j(\theta_{ij})\otimes I_{\overline{j}}^{n-1}\right\}=\left\{RX_j(\theta_{ij}), RY_j(\theta_{ij}), RZ_j(\theta_{ij})\right\}\otimes I_{\overline{j}}^{n-1},
\end{align}
where \(\overline{j}\) means except for position j, \( i, j \) represent the \(j\)-th qubit at the \(i\)-th layer of depth.
Using the same method as in the previous subsection, we can compute \( \mathrm{E}\left[ U(\theta_{ij})^\dagger A U(\theta_{ij}) \right] \) for the \( 2 \)-qubit case under two scenarios:
\begin{align}
    \label{eq: u_1}
    U_1 &\in \left\{RX(\theta)\otimes I, RY(\theta)\otimes I, RZ(\theta)\otimes I\right\} = U\otimes I,\\
    \label{eq: u_2}
    U_2 &\in \left\{I \otimes RX(\theta), I\otimes RY(\theta), I \otimes RZ(\theta) \right\}= I \otimes U.
\end{align}

Let's calculate with the case of $U_1$, and we need first definite the form of matrix $A$ as follows:
\begin{align}
    A = \begin{bmatrix}
        a_{11} & a_{12} & a_{13} & a_{14} \\
        a_{21} & a_{22} & a_{23} & a_{24} \\
        a_{31} & a_{32} & a_{33} & a_{34} \\
        a_{41} & a_{42} & a_{43} & a_{44}
    \end{bmatrix}.
\end{align}
\begin{enumerate}
    \item For \(RX(\theta)\otimes I\):
    \begin{align}
        &\mathrm{E}\left[ \left(RX(\theta)\otimes I\right)^\dagger \cdot A\cdot \left(RX(\theta)\otimes I\right) \right]\\
        =&\mathrm{E}\left[
\begin{bmatrix}
\cos\frac{\theta}{2} & 0 & -i\sin\frac{\theta}{2} & 0 \\
0 & \cos\frac{\theta}{2} & 0 & -i\sin\frac{\theta}{2} \\
-i\sin\frac{\theta}{2} & 0 & \cos\frac{\theta}{2} & 0 \\
0 & -i\sin\frac{\theta}{2} & 0 & \cos\frac{\theta}{2}
\end{bmatrix}^\dagger \cdot \begin{bmatrix}
        a_{11} & a_{12} & a_{13} & a_{14} \\
        a_{21} & a_{22} & a_{23} & a_{24} \\
        a_{31} & a_{32} & a_{33} & a_{34} \\
        a_{41} & a_{42} & a_{43} & a_{44}
    \end{bmatrix} \cdot \begin{bmatrix}
\cos\frac{\theta}{2} & 0 & -i\sin\frac{\theta}{2} & 0 \\
0 & \cos\frac{\theta}{2} & 0 & -i\sin\frac{\theta}{2} \\
-i\sin\frac{\theta}{2} & 0 & \cos\frac{\theta}{2} & 0 \\
0 & -i\sin\frac{\theta}{2} & 0 & \cos\frac{\theta}{2}
\end{bmatrix}\right]\\
\label{eq: rxi}
=& \frac{1}{2} \begin{bmatrix}
a_{11} + a_{33} & a_{12} + a_{34} & a_{13} + a_{31} & a_{14} + a_{32} \\
a_{21} + a_{43} & a_{22} + a_{44} & a_{23} + a_{41} & a_{24} + a_{42} \\
a_{31} + a_{13} & a_{32} + a_{14} & a_{33} + a_{11} & a_{34} + a_{12} \\
a_{41} + a_{23} & a_{42} + a_{24} & a_{43} + a_{21} & a_{44} + a_{22}
\end{bmatrix}.
    \end{align}
    
    \item For \(RY(\theta)\otimes I\):
       \begin{align}
        &\mathrm{E}\left[ \left(RY(\theta)\otimes I\right)^\dagger \cdot A\cdot \left(RY(\theta)\otimes I\right) \right]\\
        =&\mathrm{E}\left[
\begin{bmatrix}
\cos\frac{\theta}{2} & 0 & -\sin\frac{\theta}{2} & 0 \\
0 & \cos\frac{\theta}{2} & 0 & -\sin\frac{\theta}{2} \\
\sin\frac{\theta}{2} & 0 & \cos\frac{\theta}{2} & 0 \\
0 & \sin\frac{\theta}{2} & 0 & \cos\frac{\theta}{2}
\end{bmatrix}^\dagger \cdot \begin{bmatrix}
        a_{11} & a_{12} & a_{13} & a_{14} \\
        a_{21} & a_{22} & a_{23} & a_{24} \\
        a_{31} & a_{32} & a_{33} & a_{34} \\
        a_{41} & a_{42} & a_{43} & a_{44}
    \end{bmatrix} \cdot \begin{bmatrix}
\cos\frac{\theta}{2} & 0 & -\sin\frac{\theta}{2} & 0 \\
0 & \cos\frac{\theta}{2} & 0 & -\sin\frac{\theta}{2} \\
\sin\frac{\theta}{2} & 0 & \cos\frac{\theta}{2} & 0 \\
0 & \sin\frac{\theta}{2} & 0 & \cos\frac{\theta}{2}
\end{bmatrix}\right]\\
\label{eq: ryi}
=&\frac{1}{2} \begin{bmatrix}
a_{11} + a_{33} & a_{12} + a_{34} & a_{13} - a_{31} & a_{14} - a_{32} \\
a_{21} + a_{43} & a_{22} + a_{44} & a_{23} - a_{41} & a_{24} - a_{42} \\
a_{31} - a_{13} & a_{32} - a_{14} & a_{33} + a_{11} & a_{34} + a_{12} \\
a_{41} - a_{23} & a_{42} - a_{24} & a_{43} + a_{21} & a_{44} + a_{22}
\end{bmatrix}.
    \end{align}

    \item For \(RZ(\theta)\otimes I\):
           \begin{align}
        &\mathrm{E}\left[ \left(RZ(\theta)\otimes I\right)^\dagger \cdot A\cdot \left(RZ(\theta)\otimes I\right) \right]\\
        =&\mathrm{E}\left[
\begin{bmatrix}
1 & 0 & 0 & 0 \\
0 & 1 & 0 & 0 \\
0 & 0 & e^{i\theta} & 0 \\
0 & 0 & 0 & e^{i\theta}
\end{bmatrix}^\dagger \cdot \begin{bmatrix}
        a_{11} & a_{12} & a_{13} & a_{14} \\
        a_{21} & a_{22} & a_{23} & a_{24} \\
        a_{31} & a_{32} & a_{33} & a_{34} \\
        a_{41} & a_{42} & a_{43} & a_{44}
    \end{bmatrix} \cdot \begin{bmatrix}
1 & 0 & 0 & 0 \\
0 & 1 & 0 & 0 \\
0 & 0 & e^{i\theta} & 0 \\
0 & 0 & 0 & e^{i\theta}
\end{bmatrix}\right]\\
\label{eq: rzi}
=&\mathrm{E}\left[\begin{matrix}a_{11} & a_{12} & a_{13} e^{i\theta} & a_{14} e^{i\theta}\\a_{21} & a_{22} & a_{23} e^{i\theta} & a_{24} e^{i\theta}\\a_{31} e^{- i\theta} & a_{32} e^{- i\theta} & a_{33} & a_{34}\\a_{41} e^{- i\theta} & a_{42} e^{- i\theta} & a_{43} & a_{44}\end{matrix}\right]= \begin{bmatrix}
a_{11} & a_{12} & & \\
a_{21} & a_{22} & & \\
& & a_{33} & a_{34} \\
& & a_{43} & a_{44}
\end{bmatrix}.
    \end{align}
    
\end{enumerate}
Eq.~\eqref{eq: rxi} and Eq.~\eqref{eq: ryi} have used the conclusion of Eq.\eqref{eq: two}, and Eq.~\eqref{eq: rzi} uses the result of Eq.\eqref{eq: exp}. 

Upon these results, we can get the result of \( \mathrm{E}\left[ U_1(\theta)^\dagger A U_1(\theta) \right] \) under \(U_1\in U\otimes I\) as follows:
\begin{align}
    &\mathrm{E}\left[ U_1(\theta)^\dagger A U_1(\theta) \right]\\
    = &\frac{1}{3}\left\{\mathrm{E}\left[ \left(RX(\theta)\otimes I\right)^\dagger A\left(RX(\theta)\otimes I\right) \right]+\mathrm{E}\left[ \left(RY(\theta)\otimes I\right)^\dagger  A\left(RY(\theta)\otimes I\right) \right]+\mathrm{E}\left[ \left(RZ(\theta)\otimes I\right)^\dagger A\left(RZ(\theta)\otimes I\right) \right]\right\}\\
    \label{eq: ui}
    =& \frac{1}{3} \begin{bmatrix}
2a_{11} + a_{33} & 2a_{12} + a_{34} & a_{13} & a_{14} \\
2a_{21} + a_{43} & 2a_{22} + a_{44} & a_{23} & a_{24} \\
a_{31} & a_{32} & 2a_{33} + a_{11} & 2a_{34} + a_{12} \\
a_{41} & a_{42} & 2a_{43} + a_{21} & 2a_{44} + a_{22}
\end{bmatrix}\\
=& \frac{1}{3}\left\{ A+ \begin{bmatrix}
a_{11} + a_{33} & a_{12} + a_{34} & & \\
a_{21} + a_{43} & a_{22} + a_{44} & & \\
& & a_{33} + a_{11} & a_{34} + a_{12} \\
& & a_{43} + a_{21} & a_{44} + a_{22}
\end{bmatrix}\right\} = \frac{1}{3}\left\{A + I\otimes \text{Tr}_U A\right\}.
\end{align}

As the same, then we calculate with the case of $U_2$.

\begin{enumerate}
    \item For \(I \otimes RX(\theta)\):
\begin{align}
        &\mathrm{E}\left[ \left(I \otimes RX(\theta)\right)^\dagger \cdot A\cdot \left(I \otimes RX(\theta)\right) \right]\\
        =&\mathrm{E}\left[
\begin{bmatrix}
\cos\frac{\theta}{2} & -i\sin\frac{\theta}{2} & 0 & 0 \\
-i\sin\frac{\theta}{2} & \cos\frac{\theta}{2} & 0 & 0 \\
0 & 0 & \cos\frac{\theta}{2} & -i\sin\frac{\theta}{2} \\
0 & 0 & -i\sin\frac{\theta}{2} & \cos\frac{\theta}{2}
\end{bmatrix}^\dagger \cdot \begin{bmatrix}
        a_{11} & a_{12} & a_{13} & a_{14} \\
        a_{21} & a_{22} & a_{23} & a_{24} \\
        a_{31} & a_{32} & a_{33} & a_{34} \\
        a_{41} & a_{42} & a_{43} & a_{44}
    \end{bmatrix} \cdot \begin{bmatrix}
\cos\frac{\theta}{2} & -i\sin\frac{\theta}{2} & 0 & 0 \\
-i\sin\frac{\theta}{2} & \cos\frac{\theta}{2} & 0 & 0 \\
0 & 0 & \cos\frac{\theta}{2} & -i\sin\frac{\theta}{2} \\
0 & 0 & -i\sin\frac{\theta}{2} & \cos\frac{\theta}{2}
\end{bmatrix}\right]\\
\label{eq: irx}
=&\frac{1}{2}\begin{bmatrix}
a_{11} + a_{22} & a_{12} + a_{21} & a_{13} + a_{24} & a_{14} + a_{23} \\
a_{21} + a_{12} & a_{22} + a_{11} & a_{23} + a_{14} & a_{24} + a_{13} \\
a_{31} + a_{42} & a_{32} + a_{41} & a_{33} + a_{44} & a_{34} + a_{43} \\
a_{41} + a_{32} & a_{42} + a_{31} & a_{43} + a_{34} & a_{44} + a_{33}
\end{bmatrix}.
    \end{align}

    \item For \(I \otimes RY(\theta)\):
\begin{align}
        &\mathrm{E}\left[ \left(I \otimes RY(\theta)\right)^\dagger \cdot A\cdot \left(I \otimes RY(\theta)\right) \right]\\
        =&\mathrm{E}\left[\begin{bmatrix}
\cos\frac{\theta}{2} & -\sin\frac{\theta}{2} & 0 & 0 \\
\sin\frac{\theta}{2} & \cos\frac{\theta}{2} & 0 & 0 \\
0 & 0 & \cos\frac{\theta}{2} & -\sin\frac{\theta}{2} \\
0 & 0 & \sin\frac{\theta}{2} & \cos\frac{\theta}{2}
\end{bmatrix}^\dagger \cdot \begin{bmatrix}
        a_{11} & a_{12} & a_{13} & a_{14} \\
        a_{21} & a_{22} & a_{23} & a_{24} \\
        a_{31} & a_{32} & a_{33} & a_{34} \\
        a_{41} & a_{42} & a_{43} & a_{44}
    \end{bmatrix} \cdot \begin{bmatrix}
\cos\frac{\theta}{2} & -\sin\frac{\theta}{2} & 0 & 0 \\
\sin\frac{\theta}{2} & \cos\frac{\theta}{2} & 0 & 0 \\
0 & 0 & \cos\frac{\theta}{2} & -\sin\frac{\theta}{2} \\
0 & 0 & \sin\frac{\theta}{2} & \cos\frac{\theta}{2}
\end{bmatrix}\right]\\
\label{eq: iry}
=&\frac{1}{2}\begin{bmatrix}
a_{11} + a_{22} & a_{12} - a_{21} & a_{13} + a_{24} & a_{14} - a_{23} \\
a_{21} - a_{12} & a_{22} + a_{11} & a_{23} - a_{14} & a_{24} + a_{13} \\
a_{31} + a_{42} & a_{32} - a_{41} & a_{33} + a_{44} & a_{34} - a_{43} \\
a_{41} - a_{32} & a_{42} + a_{31} & a_{43} - a_{34} & a_{44} + a_{33}
\end{bmatrix}.
    \end{align}

    \item \(I \otimes RZ(\theta)\):
\begin{align}
        &\mathrm{E}\left[ \left(I \otimes RZ(\theta)\right)^\dagger \cdot A\cdot \left(I \otimes RZ(\theta)\right) \right]\\
        =&\mathrm{E}\left[\begin{bmatrix}
1 & 0 & 0 & 0 \\
0 & e^{i\theta} & 0 & 0 \\
0 & 0 & 1 & 0 \\
0 & 0 & 0 & e^{i\theta}
\end{bmatrix}^\dagger \cdot \begin{bmatrix}
        a_{11} & a_{12} & a_{13} & a_{14} \\
        a_{21} & a_{22} & a_{23} & a_{24} \\
        a_{31} & a_{32} & a_{33} & a_{34} \\
        a_{41} & a_{42} & a_{43} & a_{44}
    \end{bmatrix} \cdot \begin{bmatrix}
1 & 0 & 0 & 0 \\
0 & e^{i\theta} & 0 & 0 \\
0 & 0 & 1 & 0 \\
0 & 0 & 0 & e^{i\theta}
\end{bmatrix}\right]\\
\label{eq: irz}
=&\left[\begin{matrix}a_{11} & a_{12} e^{i\theta} & a_{13} & a_{14} e^{i\theta}\\a_{21} e^{- i\theta} & a_{22} & a_{23} e^{- i\theta} & a_{24}\\a_{31} & a_{32} e^{i\theta} & a_{33} & a_{34} e^{i\theta}\\o_{41} e^{- i\theta} & a_{42} & a_{43} e^{- i\theta} & a_{44}\end{matrix}\right] = \begin{bmatrix}
a_{11} & & a_{13} & \\
& a_{22} & & a_{24} \\
a_{31}& & a_{33} & \\
& a_{42} & & a_{44}
\end{bmatrix}.
\end{align}
    
\end{enumerate}
Eq.~\eqref{eq: irx} and Eq.~\eqref{eq: iry} have used the conclusion of Eq.\eqref{eq: two}, and Eq.~\eqref{eq: irz} uses the result of Eq.\eqref{eq: exp}. 

Then, we can get the result of \( \mathrm{E}\left[ U_2(\theta)^\dagger A U_2(\theta) \right] \) under \(U_2\in I\otimes U\) as follows:
\begin{align}
    &\mathrm{E}\left[ U_2(\theta)^\dagger A U_2(\theta) \right]\\
    = &\frac{1}{3}\left\{\mathrm{E}\left[ \left(I\otimes RX(\theta)\right)^\dagger A\left(I\otimes RX(\theta)\right) \right]+\mathrm{E}\left[ \left(I\otimes RY(\theta)\right)^\dagger A\left(I\otimes RY(\theta)\right) \right]+\mathrm{E}\left[ \left(I\otimes RZ(\theta)\right)^\dagger A\left(I\otimes RZ(\theta)\right) \right]\right\}\\
    \label{eq: iu}
    = &\frac{1}{3}\begin{bmatrix}
2a_{11} + a_{22} & a_{12} & 2a_{13} + a_{24} & a_{14} \\
a_{21} & 2a_{22} + a_{11} & a_{23} & 2a_{24} + a_{13} \\
2a_{31} + a_{42} & a_{32} & 2a_{33} + a_{44} & a_{34} \\
a_{41} & 2a_{42} + a_{31} & a_{43} & 2a_{44} + a_{33}
\end{bmatrix}\\
=& \frac{1}{3}\left\{ A+ \begin{bmatrix}
a_{11} + a_{22} & & a_{13} + a_{24} & \\
& a_{22} + a_{11} & & a_{24} + a_{13} \\
a_{31} + a_{42} & & a_{33} + a_{44} & \\
& a_{42} + a_{31} & & a_{44} + a_{33}
\end{bmatrix}\right\} = \frac{1}{3}\left\{A + \text{Tr}_U A \otimes I\right\}.
\end{align}

Building upon the conclusions from the single-qubit and \( 2 \)-qubit scenarios as shown in the Eq.~\eqref{eq: u}, Eq.~\eqref{eq: iu}, and Eq.~\eqref{eq: ui}, we can extend the computation to \( 3 \)-qubit and larger \( n \)-qubit systems. Although the process becomes increasingly complex and computationally intensive, the final forms of \( \mathrm{E}_{ij}\left[ U(\theta_{ij})^\dagger A U(\theta_{ij}) \right] \) can be unified as:
\begin{align}
    \mathrm{E}\left[ U(\theta_{ij})^\dagger A U(\theta_{ij}) \right]=\frac{1}{3}\left\{A+I_j\otimes\text{Tr}_j A\right\}.
\end{align}

\section{Derivation of Eq.~\UAUall{}\label{app: 24}}
With the results above, we can begin analysis the $n$-qubit, single-depth, and $n$-parameter \( \mathrm{E}\left[ U(\theta_{\mathbf{i}})^\dagger A U(\theta_{\mathbf{i}}) \right] \), where \(\mathbf{i}\) represents the $i$-th layer of depth, and \(\theta_{\mathbf{i}}\) is a list of parameters with length $n$.
First, the unitary operator can be decomposed as follows:
\begin{align}
    U_{i}(\theta_{i}) \in \left\{\bigotimes_{j=1}^{n} \left\{RX(\theta_{ij}),RY(\theta_{ij}),RZ(\theta_{ij})\right\} \right\} = \left\{\prod_{j=1}^{n}\left\{RX(\theta_{ij}),RY(\theta_{ij}),RZ(\theta_{ij})\right\}\otimes I_{\overline{j}}^{n-1}\right\}.
\end{align}
So, we can use the following equation to decompose the \( n \)-parameter expectation into \( n \) single-parameter \( \mathrm{E}\left[ U(\theta_{ij})^\dagger A U(\theta_{ij}) \right] \), we can generalize the result for any positive integer \( n \) as: 
\begin{align}
    \mathrm{E}_{\mathbf{i}}\left[ U(\theta_{\mathbf{i}})^\dagger A U(\theta_{\mathbf{i}}) \right] =& E_{(i1, i2, \dots, in)}\left[ \big(U(\theta_{\mathbf{i1}})U(\theta_{\mathbf{i2}})\cdots U(\theta_{\mathbf{in}}) \big)^\dagger \cdot A \cdot \big(U(\theta_{\mathbf{i1}})U(\theta_{\mathbf{i2}})\cdots U(\theta_{\mathbf{in}}) \big)\right]\\=& \frac{1}{3^n}\sum_{\sigma \in \mathcal{P}(n)}\left\{I_\sigma^{\otimes|\sigma|}\otimes\text{Tr}_\sigma A \right\},
\end{align}
where \(\mathcal{P}(n)\) is the power set of \(\{1,2,\cdots,n\}\).

To illustrate this process more intuitively, we provide examples for \( n = 2 \) and \( n = 3 \):  
\begin{enumerate}
    \item When \( n = 2 \), the power set is \(\mathcal{P}(2)=\mathcal{P}(\{j, j+1\}) = \{\varnothing, \{j\}, \{j+1\}, \{j,j+1\}\}\):
    \begin{align}
        \mathrm{E} \left[ U_{i(j+1)}^\dagger U_{ij}^\dagger\cdot A \cdot U_{ij}U_{i(j+1)} \right] &= \frac{1}{3}\mathrm{E} \left[ U_{i(j+1)}^\dagger\left\{A+I_j\otimes\text{Tr}_j A \right\}U_{i(j+1)} \right]\\
        & = \frac{1}{9}\left\{\left\{A+I_j\otimes\text{Tr}_j A \right\} + I_{j+1}\otimes \left\{A+I_j\otimes\text{Tr}_j A \right\} \right\}\\
        & = \frac{1}{9}\left\{A + I_{j} \otimes \text{Tr}_j A + I_{j+1} \otimes \text{Tr}_{j+1} A + I_{j} \otimes I_{j+1} \otimes \text{Tr}_{j,j+1} A\right\}\\
        & = \frac{1}{3^2}\sum_{\mathcal{P}(2)}\left\{I_\sigma^{\otimes|\sigma|}\otimes\text{Tr}_\sigma A \right\}.
    \end{align}

    \item Similarly, when \( n = 3 \), the power set becomes 
    \begin{align}
    \mathcal{P}(3)=\mathcal{P}(\{j, j+1, j+2\}) = \{\varnothing, \{j\}, \{j+1\}, \{j+2\}, \{j,j+1\}, \{j,j+2\}, \{j+1,j+2\}, \{j,j+1,j+2\}\}.
    \end{align}
    The expectation can be shown as:
    \begin{align}
        & \mathrm{E} \left[U_{i(j+2)}^\dagger U_{i(j+1)}^\dagger U_{ij}^\dagger\cdot A \cdot U_{ij}U_{i(j+1)}U_{i(j+2)} \right]\\
        = & \frac{1}{9}\mathrm{E} \left[ U_{i(j+1)}^\dagger\left\{A + I_{j} \otimes \text{Tr}_j A + I_{j+1} \otimes \text{Tr}_{j+1} A + I_{j} \otimes I_{j+1} \otimes \text{Tr}_{j,j+1} A \right\}U_{i(j+1)} \right] \\
        = & \frac{1}{27} \{ A + I_{j} \otimes \text{Tr}_j A + I_{j+1} \otimes \text{Tr}_{j+1} A + I_{j+2} \otimes \text{Tr}_{j+2} A + I_{j} \otimes I_{j+1} \otimes \text{Tr}_{j,j+1} A\\ &+ I_{j} \otimes I_{j+2} \otimes \text{Tr}_{j,j+2} A+ I_{j+1} \otimes I_{j+2} \otimes \text{Tr}_{j+1,j+1} A+ I_{j} \otimes I_{j+1} \otimes I_{j+2} \otimes \text{Tr}_{j,j+1,j+2} A\}.\notag\\
        =& \frac{1}{3^3}\sum_{\mathcal{P}(3)}\left\{I_\sigma^{\otimes|\sigma|}\otimes\text{Tr}_\sigma A \right\}.
    \end{align}
\end{enumerate}

Finally, we can analyze the circuit of $n$-qubit, $d$-depth, and $nd$-parameter by extending the depth from 1 to \( d \).
This extension builds upon the derivation from the previous subsection, and its final form is:  
\begin{align}
    \mathrm{E} \left[ U^\dagger A  U\right] = \frac{1}{3^n}\sum_{\sigma \in \mathcal{P}(n)}\left\{\left(\frac{4^{|\sigma|}}{3^n}\right)^{d-1}\cdot I_\sigma^{\otimes|\sigma|}\otimes\text{Tr}_\sigma \left\{A\right\}\right\}.
\end{align}

We observe that as the depth \( d \) increases, most terms rapidly decay due to \(\left(\frac{4^{|\sigma|}}{3^n}\right)^{d-1}\) in the equation.
This indicates that quantum coherence, represented by off-diagonal elements of the system, is gradually washed out, leading to a transition toward a classical distribution state.  

To aid in understanding this calculation process, we provide examples for \( n = 2 \) and \( n = 3 \) with depth \( d \).
\begin{enumerate}
\item When \( n = 2 \) and \( d = 2 \), we can get the form as following:
\begin{align}
    \mathrm{E}\left[ U^\dagger A  U\right] =&  \mathrm{E}_{\mathbf{i},\mathbf{i+1}}\left[ U(\theta_{\mathbf{i+1}})^\dagger \left\{ U(\theta_{\mathbf{i}})^\dagger A U(\theta_{\mathbf{i}}) \right\} U(\theta_{\mathbf{i+1}}) \right]\\
    =& \frac{1}{3^2}\mathrm{E}_{\mathbf{i+1}}\left[ U(\theta_{\mathbf{i+1}})^\dagger \left\{ \sum_{\mathcal{P}(2)}\left\{I_\sigma^{\otimes|\sigma|}\otimes\text{Tr}_\sigma A \right\} \right\} U(\theta_{\mathbf{i+1}}) \right]\\
    =& \frac{1}{3^2}\sum_{\mathcal{P}(2)}\left\{I_\sigma^{\otimes|\sigma|}\otimes\text{Tr}_\sigma \left\{\frac{1}{3^2}\sum_{\mathcal{P}(2)}\left\{I_\sigma^{\otimes|\sigma|}\otimes\text{Tr}_\sigma A \right\}\right\} \right\}\\
    =& \frac{1}{3^2}\Bigg\{\frac{1}{3^2} \sum_{\mathcal{P}(2)}\left\{I_\sigma^{\otimes|\sigma|}\otimes\text{Tr}_\sigma A \right\}+\frac{1}{3^2}I_{j}\otimes\Tr_{j}\left(\sum_{\mathcal{P}(2)}\left\{I_\sigma^{\otimes|\sigma|}\otimes\text{Tr}_\sigma A \right\}\right)\\&+\frac{1}{3^2}I_{j+1}\otimes\Tr_{j+1}\left(\sum_{\mathcal{P}(2)}\left\{I_\sigma^{\otimes|\sigma|}\otimes\text{Tr}_\sigma A \right\}\right)+\frac{1}{3^2}I\cdot\Tr\left(\sum_{\mathcal{P}(2)}\left\{I_\sigma^{\otimes|\sigma|}\otimes\text{Tr}_\sigma A \right\}\right)\Bigg\}\notag \\
    = & \frac{1}{3^2}\bigg\{\frac{1}{3^2} \left\{A + I_{j} \otimes \text{Tr}_j A + I_{j+1} \otimes \text{Tr}_{j+1} A + I_{j} \otimes I_{j+1} \otimes \text{Tr}_{j,j+1} A\right\}\\
    &+ \frac{1}{3^2} I_j\otimes \Tr_{j}\left\{A + I_{j} \otimes \text{Tr}_j A + I_{j+1} \otimes \text{Tr}_{j+1} A + I_{j} \otimes I_{j+1} \otimes \text{Tr}_{j,j+1} A\right\}\notag\\
    &+ \frac{1}{3^2} I_{j+1}\otimes \Tr_{j+1}\left\{A + I_{j} \otimes \text{Tr}_j A + I_{j+1} \otimes \text{Tr}_{j+1} A + I_{j} \otimes I_{j+1} \otimes \text{Tr}_{j,j+1} A\right\}\notag\\
    &+ \frac{1}{3^2} I\cdot \Tr\left\{A + I_{j} \otimes \text{Tr}_j A + I_{j+1} \otimes \text{Tr}_{j+1} A + I_{j} \otimes I_{j+1} \otimes \text{Tr}_{j,j+1} A\right\}\bigg\}\notag \\
    =&\frac{1}{3^2}\left\{\frac{1}{3^2}A+\frac{4}{3^2}I_{j}\otimes\Tr_{j}A+\frac{4}{3^2}I_{j+1}\otimes\Tr_{j+1}A+\frac{16}{3^2}I\cdot \Tr A\right\}\\
    =&\frac{1}{3^2}\sum_{\sigma \in \mathcal{P}(2)}\left\{\left(\frac{4^{|\sigma|}}{3^2}\right)\cdot I_\sigma^{\otimes|\sigma|}\otimes\text{Tr}_\sigma \left\{A\right\}\right\}.
\end{align}

For \( n = 2 \) and \( d = 3 \), we can use the conclusion straightly.
\begin{align}
        \mathrm{E}\left[ U^\dagger A  U\right] =&  \mathrm{E}_{\mathbf{i},\mathbf{i+1},\mathbf{i+2}}\left[ U(\theta_{\mathbf{i+2}})^\dagger \left\{ U(\theta_{\mathbf{i+1}})^\dagger U(\theta_{\mathbf{i}})^\dagger A U(\theta_{\mathbf{i}}) U(\theta_{\mathbf{i+1}}) \right\} U(\theta_{\mathbf{i+2}}) \right]\\
        =& \frac{1}{3^2}\mathrm{E}_{\mathbf{i+2}}\left[ U(\theta_{\mathbf{i+2}})^\dagger \left\{ \sum_{\mathcal{P}(2)}\left(\frac{4^{|\sigma|}}{3^2}\right)\left\{I_\sigma^{\otimes|\sigma|}\otimes\text{Tr}_\sigma A \right\} \right\} U(\theta_{\mathbf{i+2}}) \right]\\
        =&\frac{1}{3^2}\left\{\frac{1}{3^4}A+\frac{4^2}{3^4}I_{j}\otimes\Tr_{j}A+\frac{4^2}{3^4}I_{j+1}\otimes\Tr_{j+1}A+\frac{16^2}{3^4}I\cdot \Tr A\right\}\\
        =&\frac{1}{3^2}\sum_{\sigma \in \mathcal{P}(2)}\left\{\left(\frac{4^{|\sigma|}}{3^2}\right)^2\cdot I_\sigma^{\otimes|\sigma|}\otimes\text{Tr}_\sigma \left\{A\right\}\right\}.
\end{align}

For more depth, it will only influence the coefficient in this equation. We can summarize the term of \(\mathrm{E}\left[ U^\dagger A  U\right]\).

\item When we consider \( n = 3 \) and \( d = 2 \), it will be more complex, but we can get the same result:
\begin{align}
    \mathrm{E}\left[ U^\dagger A  U\right] =&  \mathrm{E}_{\mathbf{i},\mathbf{i+1},\mathbf{i+2}}\left[ U(\theta_{\mathbf{i+2}})^\dagger \left\{ U(\theta_{\mathbf{i+1}})^\dagger U(\theta_{\mathbf{i}})^\dagger A U(\theta_{\mathbf{i}}) U(\theta_{\mathbf{i+1}}) \right\} U(\theta_{\mathbf{i+2}}) \right]\\
    =& \frac{1}{3^3} \{ \frac{1}{3^3}A + \frac{4}{3^3} I_{j} \otimes \text{Tr}_j A + \frac{4}{3^3}I_{j+1} \otimes \text{Tr}_{j+1} A + \frac{4}{3^3}I_{j+2} \otimes \text{Tr}_{j+2} A + \frac{16}{3^3}I_{j} \otimes I_{j+1} \otimes \text{Tr}_{j,j+1} A\\ &+ \frac{16}{3^3}I_{j} \otimes I_{j+2} \otimes \text{Tr}_{j,j+2} A+ \frac{16}{3^3}I_{j+1} \otimes I_{j+2} \otimes \text{Tr}_{j+1,j+1} A+ \frac{64}{3^3}I \cdot \text{Tr} A\}.\notag\\
    =& \frac{1}{3^3}\sum_{\sigma \in \mathcal{P}(3)}\left\{\left(\frac{4^{|\sigma|}}{3^2}\right)\cdot I_\sigma^{\otimes|\sigma|}\otimes\text{Tr}_\sigma \left\{A\right\}\right\}.
\end{align}

For \( n = 3 \) and \( d = 3 \), we can also get the same result as the following:
\begin{align}
     \mathrm{E}\left[ U^\dagger A  U\right] =&  \mathrm{E}_{\mathbf{i},\mathbf{i+1},\mathbf{i+2}}\left[ U(\theta_{\mathbf{i+2}})^\dagger \left\{ U(\theta_{\mathbf{i+1}})^\dagger U(\theta_{\mathbf{i}})^\dagger A U(\theta_{\mathbf{i}}) U(\theta_{\mathbf{i+1}}) \right\} U(\theta_{\mathbf{i+2}}) \right]\\
     =& \frac{1}{3^3} \{ \frac{1}{3^3}A + \frac{4^2}{3^3} I_{j} \otimes \text{Tr}_j A + \frac{4^2}{3^3}I_{j+1} \otimes \text{Tr}_{j+1} A + \frac{4^2}{3^3}I_{j+2} \otimes \text{Tr}_{j+2} A + \frac{16^2}{3^3}I_{j} \otimes I_{j+1} \otimes \text{Tr}_{j,j+1} A\\ &+ \frac{16^2}{3^3}I_{j} \otimes I_{j+2} \otimes \text{Tr}_{j,j+2} A+ \frac{16^2}{3^3}I_{j+1} \otimes I_{j+2} \otimes \text{Tr}_{j+1,j+1} A+ \frac{64^2}{3^3}I \cdot \text{Tr} A\}.\notag\\
     \label{eq: final_fomula}
    =& \frac{1}{3^3}\sum_{\sigma \in \mathcal{P}(3)}\left\{\left(\frac{4^{|\sigma|}}{3^2}\right)^2\cdot I_\sigma^{\otimes|\sigma|}\otimes\text{Tr}_\sigma \left\{A\right\}\right\}.
\end{align}

For more depth, the result is the same as the situation of \(n=2\) and can be summed up as \(\mathrm{E}\left[ U^\dagger A  U\right]\).
\end{enumerate}

These examples demonstrate how depth influences this equation.

\section{Derivation of Eq.~\UAUBUCUpoint{}\label{app: 32}}

Similar to the computation of \(\mathrm{E}\left[ U^\dagger A U \right] \), we start with the simplest case: a single-qubit, single-depth, and single-parameter \( \mathrm{E}\left[ U^\dagger A UBU^\dagger C U \right] \).
However, we observe that \(\mathrm{E}\left[ U^\dagger A UBU^\dagger C U \right] \) cannot be perfectly expressed only using \( A, B, C \) as \(\mathrm{E}\left[ U^\dagger A U\right] \) can.
As a result, during the calculation, we need to discard some smaller terms to simplify the expression.

The subsequent steps mirror those of \(\mathrm{E}\left[ U^\dagger A U \right] \): we first extend to \( n \)-qubit, single-depth, single-parameter \(\mathrm{E}\left[ U^\dagger A UBU^\dagger C U \right] \).
Next, generalize to \( n \)-qubit, single-depth, \(n\)-parameter result.
Finally, construct the complete model for \( n \)-qubit, \( d \)-depth, and \( nd \)-parameter systems.

Although our ultimate goal is to compute \( \mathrm{E}\left[ U^\dagger AUBU^\dagger A U \right] \), all discussions here are based on \( \mathrm{E}\left[ U^\dagger A UBU^\dagger C U \right] \) to avoid results like \( a_{ij}b_{jk}a_{kl} \) that we can not find the element $a$ is from which matrix.  

Even for a 1-qubit, 1-depth system, by using the Eq.\eqref{eq: one}, Eq.\eqref{eq: two}, Eq.\eqref{eq: three}, Eq.\eqref{eq: four}, Eq.\eqref{eq: exp} each entry in the matrix is composed of $64$ polynomials, making it nearly impossible to decompose the matrix directly. However, we observe that when \( A, B, C \) are set to the identity matrix \( I \), we can derive the following conclusion:
\begin{align}
    \text{When}\, A = I:  &\mathrm{E}\left[ U^\dagger AUBU^\dagger C U \right]=\mathrm{E}\left[ BU^\dagger C U \right]=B\cdot\{C+I\cdot\Tr{C}\}=BC+\Tr{C}B, \\
    \text{When}\, B = I:  &\mathrm{E}\left[ U^\dagger AUBU^\dagger C U \right]=\mathrm{E}\left[ U^\dagger AC U \right]=AC+\Tr{AC},\\ 
    \text{When}\, C = I:  &\mathrm{E}\left[ U^\dagger AUBU^\dagger C U \right]=\mathrm{E}\left[ U^\dagger A U B\right]=\{A+I\cdot\Tr{A}\}\cdot B=AB+\Tr{A}B. 
\end{align}

Based on this result, we identify two operational combinations that satisfy this conclusion:
\begin{align}
    \label{eq: fABC}
    f &= ABC+\text{Tr}\{AC\}B,\\
    \label{eq: gABC}
    g &= \text{Tr}\{ABC\}\cdot I-\text{Tr}\{BC\}\cdot A -\text{Tr}\{AB\}\cdot C + \text{Tr}\{A\}\cdot BC+\text{Tr}\{B\}\cdot AC+\text{Tr}\{C\}\cdot AB-ABC.
\end{align}

Through our tests, we find that \( \mathrm{E}\left[ U^\dagger AUBU^\dagger C U \right] \) can be decomposed into the following structure:
\begin{align}
    \mathrm{E}\left[ U^\dagger AUBU^\dagger C U \right] = \frac{f+\epsilon}{4}+\frac{g}{12},
\end{align}
where \( \epsilon \) is a matrix with a trace equal to zero.
Since trace operations are frequently used in higher-dimensional systems, we can safely neglect this term for simplicity.

By examining the previous results, we can analogize the conclusions from the Supplementary.\ref{app: 22} to compute an approximation for \( \mathrm{E}\left[ U^\dagger AUBU^\dagger C U \right] \) in the case of \( n \)-qubit, single-depth, and single-parameter systems.
\begin{align}
    \mathrm{E}\left[ U(\theta_{ij})^\dagger AU(\theta_{ij})BU(\theta_{ij})^\dagger C U(\theta_{ij}) \right]= \frac{f_j}{4}+\frac{g_j}{12},
\end{align}
where \(f_j\) and \(g_j\) is related to Eq.\eqref{eq: fABC} and Eq.\eqref{eq: gABC}, respectively. The \(g_j\) can be shown as:
\begin{align}
    f_j = ABC+I_j \otimes \text{Tr}_{j}\{AC\}\cdot B.
\end{align}

When \( n \) becomes large, the coefficients of \( f_j \) in the equation are significantly larger than those of \( g_j \).
Therefore, we do not need to explicitly compute the form of \( g_j \) in this context.

\section{Derivation of Eq.~\UAUBUCUall{}\label{app: 34}}
With the results above, we can now begin analyzing the circuit of $n$-qubit, single-depth, and $n$-parameter \( \mathrm{E}_{\mathbf{i}} \left[ U(\theta_{\mathbf{i}})^\dagger A  U(\theta_{\mathbf{i}})B U(\theta_{\mathbf{i}})^\dagger CU(\theta_{\mathbf{i}})\right] \). Similar to Supplementary.\ref{app: 24}, we decompose the \( n \)-parameter form into \( n \) single-parameter \( \mathrm{E}_{ij}\left[ U(\theta_{ij})^\dagger AU(\theta_{ij})BU(\theta_{ij})^\dagger C U(\theta_{ij}) \right] \), leading to:

\begin{align}
    \label{eq: 13}
    \mathrm{E}_{\mathbf{i}} \left[ U(\theta_{\mathbf{i}})^\dagger A  U(\theta_{\mathbf{i}})B U(\theta_{\mathbf{i}})^\dagger CU(\theta_{\mathbf{i}})\right]
    = \frac{1}{4^n}\sum_{\sigma \in \mathcal{P}(n)}\left\{I_\sigma^{\otimes|\sigma|}\otimes\text{Tr}_\sigma \left\{AC\right\}\cdot B \right\} +O(4^{-n}) .
\end{align}

To understand its principle clearly, let's consider an example of \(n=2\).
\begin{align}
    &\mathrm{E}_{ij,ij+1}\left[ U_{ij+1}^\dagger U_{ij}^\dagger A U_{ij}U_{ij+1}BU_{ij+1}^\dagger U_{ij}^\dagger AU_{ij}U_{ij+1} \right]\\
    =&\frac{f_{j}(E_{j+1})}{4}+\frac{g_j(E_{j+1})}{12}=\frac{f_{j}(\frac{f_{j+1}}{4}+\frac{g_{j+1}}{12})}{4}+\frac{g_j(\frac{f_{j+1}}{4}+\frac{g_{j+1}}{12})}{12}\\
    =&\frac{f_{j}(f_{j+1})}{4^2}+\frac{f_{j}(g_{j+1})+g_j(f_{j+1})}{4\cdot 12}+\frac{g_j(g_{j+1})}{12^2}=\frac{f_{j}(f_{j+1})}{4^2}+O(4^{-2})\\
    =&\frac{1}{4^2}f_j\left( ABC+I_j\otimes \text{Tr}_{j}\{AC\} \cdot B \right)+O(4^{-2})\\
    =&\frac{1}{4^2}\left( ABC+I_j\otimes \text{Tr}_{j}\{AC\} \cdot B + I_{j+1}\otimes \text{Tr}_{j+1}\{AC\} \cdot B + I_{j,j+1}\otimes \text{Tr}_{j,j+1}\{AC\} \cdot B\right)+O(4^{-2})\\
    =&\frac{1}{4^2}\sum_{\sigma \in \mathcal{P}(2)}\left\{I_\sigma^{\otimes|\sigma|}\otimes\text{Tr}_\sigma \left\{AC\right\}\cdot B \right\} +O(4^{-2})
\end{align}

Finally, similar to Supplementary.\ref{app: 24}, we can analyze the circuit of $n$-qubit, $d$-depth, and $nd$-parameter by extending the depth from 1 to \( d \), resulting in the following form:
\begin{align}
    \mathrm{E} \left[ U^\dagger A  UB U^\dagger C  U\right] = \frac{1}{4^n}\sum_{\sigma \in \mathcal{P}(n)}\left\{(4^{|\sigma|-n})^{d-1}\cdot I_\sigma^{\otimes|\sigma|}\otimes\text{Tr}_\sigma \left\{AC\right\}\cdot B \right\} +O(4^{-n}).
\end{align}

This equation decomposes the expectation into contributions from all possible partitions of the system, with coefficients determined by the depth \( d \), qubit numbers \( n \), and the structure of \( A, B \) and \( C \).
The \( O(4^{-n}) \) term accounts for higher-order corrections, which diminish as \( n \) becomes large, reflecting the approximate nature of the system's behavior.

\section{Influence of the fixed gate}

In both \(\mathrm{E}\left[ U^\dagger A  U\right]\) and \( \mathrm{E} \left[ U^\dagger A  UB U^\dagger C  U\right] \), all the unitary gates \( U \) are composed of rotation gates \( \{RX, RY, RZ\} \). However, in the calculation of gradient expectation and variance, \( U \) contains rotation gates and fixed gates \( W \).
These two cases are never equal.

Fortunately, the trace operation is used in the calculation, and introducing \( W \), which is composed of \( CZ \)-gates and identity gates, does not change the diagonal values. This means their traces will remain equal, which can be illustrated as
\begin{align}
    \Tr{\mathrm{E}\left[ U^\dagger A  U\right]}&=\Tr{\mathrm{E}_\mathbf{1, 2,\cdots, d}\left[ (U(\theta_\mathbf{1})U(\theta_\mathbf{2})\cdots U(\theta_\mathbf{d}))^\dagger A  (U(\theta_\mathbf{1})U(\theta_\mathbf{2})\cdots U(\theta_\mathbf{d}))\right]}\\&=\Tr{\mathrm{E}_\mathbf{1, 2,\cdots, d}\left[ (U(\theta_\mathbf{1})WU(\theta_\mathbf{2})W\cdots U(\theta_\mathbf{d})W)^\dagger A  (U(\theta_\mathbf{1})WU(\theta_\mathbf{2})W\cdots U(\theta_\mathbf{d})W)\right]}
\end{align}

To prove this result, we can decompose this expression into the following two equations.
\begin{align}
    \Tr{\mathrm{E}_{\mathbf{i}}\left[ (U(\theta_{\mathbf{i}})W)^\dagger A (U(\theta_{\mathbf{i}})W) \right]} &= \Tr{W^\dagger \cdot \mathrm{E}_{\mathbf{i}}\left[ U(\theta_{\mathbf{i}})^\dagger A U(\theta_{\mathbf{i}}) \right]\cdot W}\\
    &=\sum_{i=1}^{2^n}\bra{i}W^\dagger \cdot \mathrm{E}_{\mathbf{i}}\left[ U(\theta_{\mathbf{i}})^\dagger A U(\theta_{\mathbf{i}}) \right] \cdot W\ket{i}\\
    &=(\pm 1)^2\sum_{i=1}^{2^n}\bra{i} \mathrm{E}_{\mathbf{i}}\left[ U(\theta_{\mathbf{i}})^\dagger A U(\theta_{\mathbf{i}}) \right]\ket{i}\\
    &=\sum_{i=1}^{2^n}\bra{i} \mathrm{E}_{\mathbf{i}}\left[ U(\theta_{\mathbf{i}})^\dagger A U(\theta_{\mathbf{i}}) \right]\ket{i}\\
    &=\Tr{\mathrm{E}_{\mathbf{i}}\left[ U(\theta_{\mathbf{i}})^\dagger A U(\theta_{\mathbf{i}}) \right]},
\end{align}
where \( \ket{i} \) is a vector in which the \( i \)-th element is $1$ and all other elements are $0$. Since the fixed gate \( W \) is a diagonal matrix with diagonal elements equal to \( \pm 1 \), therefore \( W\ket{i} = \pm \ket{i} \).

\begin{align}
    \Tr{\mathrm{E}_{\mathbf{i}}\left[ (WU(\theta_{\mathbf{i}}))^\dagger A (WU(\theta_{\mathbf{i}})) \right]} &= \Tr{\mathrm{E}_{\mathbf{i}}\left[ U(\theta_{\mathbf{i}})^\dagger (W^\dagger AW) U(\theta_{\mathbf{i}}) \right]}\\
    &=\Tr{\frac{1}{3^n}\sum_{\sigma \in \mathcal{P}(n)}\left\{I_\sigma^{\otimes|\sigma|}\otimes\text{Tr}_\sigma \{W^\dagger AW\} \right\}}\\
    &=\frac{1}{3^n}\sum_{\sigma \in \mathcal{P}(n)}\Tr{I_\sigma^{\otimes|\sigma|}\otimes\text{Tr}_\sigma \{W^\dagger AW\} }\\
    &=\frac{1}{3^n}\sum_{\sigma \in \mathcal{P}(n)}2^{|\sigma|}\cdot\Tr{\text{Tr}_\sigma \{W^\dagger AW\}}\\
    &=\frac{1}{3^n}\sum_{\sigma \in \mathcal{P}(n)}2^{|\sigma|}\cdot\Tr{\text{Tr}_\sigma \{A\}}\\
    &=\Tr{\frac{1}{3^n}\sum_{\sigma \in \mathcal{P}(n)}\left\{I_\sigma^{\otimes|\sigma|}\otimes\text{Tr}_\sigma \{A\} \right\}}\\
    &=\Tr{\mathrm{E}_{\mathbf{i}}\left[ U(\theta_{\mathbf{i}})^\dagger A U(\theta_{\mathbf{i}}) \right]}.
\end{align}

Therefore, it can be concluded that under the trace operation, fixed gate \( W \) does not affect the result. For \( \mathrm{E} \left[ U^\dagger A  UB U^\dagger C  U\right] \), we can get the same result by proving the following equation.

\begin{align}
    &\Tr{\mathrm{E}_{\mathbf{i}} \left[ (WU(\theta_{\mathbf{i}}))^\dagger A (WU(\theta_{\mathbf{i}}))B (WU(\theta_{\mathbf{i}}))^\dagger C(WU(\theta_{\mathbf{i}}))\right]}\\
    =&\Tr{\mathrm{E}_{\mathbf{i}} \left[U(\theta_{\mathbf{i}})^\dagger \cdot (W^\dagger AW) \cdot  U(\theta_{\mathbf{i}})B U(\theta_{\mathbf{i}})^\dagger \cdot (W^\dagger CW)\cdot U(\theta_{\mathbf{i}})\right]}\\
    \label{WAW1}
    =&\Tr{\frac{1}{4^n}\sum_{\sigma \in \mathcal{P}(n)}\left\{I_\sigma^{\otimes|\sigma|}\otimes\text{Tr}_\sigma \left\{W^\dagger ACW\right\}\cdot B\right\}}\\
    \label{WaW2}
    =&\mathrm{E}_{\mathbf{i}} \left[\Tr{(WU(\theta_{\mathbf{i}}))^\dagger A  U(\theta_{\mathbf{i}})B U(\theta_{\mathbf{i}})^\dagger C(WU(\theta_{\mathbf{i}}))}\right]\\
    =&\mathrm{E}_{\mathbf{i}} \left[\Tr{A  U(\theta_{\mathbf{i}})B U(\theta_{\mathbf{i}})^\dagger C}\right]\\
    =&\Tr{\mathrm{E}_{\mathbf{i}} \left[U(\theta_{\mathbf{i}})^\dagger A  U(\theta_{\mathbf{i}})B U(\theta_{\mathbf{i}})^\dagger CU(\theta_{\mathbf{i}})\right]}.
\end{align}
From Eq.~\eqref{WAW1} to Eq.\eqref{WaW2}, we use formula Eq.\eqref{eq: 13} to eliminate two fixed gates. Then, we leverage the properties of the trace operation to cancel out the remaining fixed gates, ultimately obtaining the final result. In \( \mathrm{E} \left[ U^\dagger A  UB U^\dagger C  U\right] \), \( WU(\theta_{\mathbf{i}}) \) and \( U(\theta_{\mathbf{i}})W \) are symmetric, and thus they lead to the same conclusion.

\section{Derivation of Eq.~\expectation{}}
The calculation of \( \mathrm{E}[\partial_k \mathcal{L}] \) can be expressed in the following form:
\begin{align}
    \mathrm{E}[\partial_k \mathcal{L}]&=\frac{i}{2}\mathrm{E}\left[\text{Tr} \left\{O_+ \cdot \left[\rho_-, P_k\right]\right\}\right]\\ &= \frac{i}{2nd}\sum_{k=1}^{nd}\text{Tr} \left\{ \mathrm{E}[O_+]\cdot \mathrm{E}\left[ [\rho_-, P_k] \right] \right\}\\
    &= \frac{i}{2nd}\sum_{k=1}^{m}\text{Tr} \left\{ \mathrm{E}[O_+]\cdot \mathrm{E}\left[ [\rho_-, P_k] \right] \right\}\\
    &= \frac{i}{2nd}\sum_{k=1}^{m}\text{Tr} \left\{ \mathrm{E}[O_+]\cdot \left(\mathrm{E}\left[\rho_-\right]\cdot P_k-P_k\cdot \mathrm{E}\left[\rho_-\right] \right) \right\}\\
    &= \frac{1}{3^n}\cdot \frac{i}{2nd}\sum_{k=1}^{m}\text{Tr} \left\{ \mathrm{E}[O_+]\cdot \sum_{\sigma \in \mathcal{P}(n')}\left\{\left(\frac{4^{|\sigma|}}{3^n}\right)^{d-1}\cdot I_\sigma^{\otimes|\sigma|}\otimes\text{Tr}_\sigma \left\{\rho P_k-P_k\rho\right\}\right\} \right\}\\
    &= \frac{1}{3^n}\cdot \frac{i}{2nd}\sum_{k=1}^{m}\text{Tr} \left\{ \mathrm{E}[O_+]\cdot \sum_{\sigma \in \mathcal{P}(n')}\left\{\left(\frac{4^{|\sigma|}}{3^n}\right)^{d-1}\cdot I_\sigma^{\otimes|\sigma|}\otimes\text{Tr}_\sigma \left[\rho, P_k\right]\right\} \right\}\\
    &\propto \frac{1}{3^n},
\end{align}
where \(m\) represents the number of effective parameters, and \(n'\) means a set without the element $k$ for \(\{1, 2,\dots,n\}/\{k\}\).
The process involves three key steps.
First, we separate the parameters for \(U_+, U_-\) and the location of $U_k$.
Then, we eliminate invalid parameters that do not contribute to the calculation and remain $m$ parts. 
Finally, we apply the conclusion from the Supplementary.\ref{app: 24} to calculate the exact value of \(E\left[ [\rho_-, P_k] \right]\).

Although the partial trace of commutator \(\Tr_\sigma[\rho, P_k]\) makes the value very small, it is not strictly zero.

\section{Derivation of Eq.~\variance{}}

Finally, we compute \(\mathrm{Var}[\partial_{k}\mathcal{L}]\), with the calculation process shown as follows:
\begin{align}
    &\mathrm{Var}[\partial_{k}\mathcal{L}]\\
    =&\mathrm{E}[(\partial_k \mathcal{L})^2]-E[\partial_k \mathcal{L}]^2\\
    =&-\frac{1}{4}\mathrm{E}\left[\text{Tr} \left\{O_+ \cdot \left[\rho_-, P_k\right]\cdot O_+ \cdot \left[\rho_-, P_k\right]\right\}\right]\\=&\frac{\alpha -1}{2}\text{Tr} \left\{\mathrm{E}\left[ O_+ \cdot \rho_-\cdot P_k \cdot O_+ \cdot \rho_-\cdot P_k\right]\right\}.
\end{align}

Since the first term of \( \mathrm{E}\left[ O_+ \cdot \rho_-\cdot P_k \cdot O_+ \cdot P_k \cdot \rho_-\right] \) diverges, we need to compute additional terms.
However, this process is highly complex.
Fortunately, from the structure, we observe that although \( \mathrm{E}\left[ O_+ \cdot \rho_-\cdot P_k \cdot O_+ \cdot P_k \cdot \rho_-\right] \) is larger than \( \mathrm{E}\left[ O_+ \cdot \rho_-\cdot P_k \cdot O_+ \cdot \rho_-\cdot P_k\right] \), there exists a proportional relationship between them.
Let us denote this relationship as:
\begin{align}
\mathrm{E}\left[ O_+ \cdot \rho_-\cdot P_k \cdot O_+ \cdot P_k \cdot \rho_-\right] = \alpha\cdot \mathrm{E}\left[ O_+ \cdot \rho_-\cdot P_k \cdot O_+ \cdot \rho_-\cdot P_k\right],
\end{align}
where \( \alpha \) is a proportionality constant that characterizes the relative scale of these two expectations.

So, we need to focus on the expectation of \(\mathrm{E}\left[ O_+ \cdot \rho_-\cdot P_k \cdot O_+ \cdot \rho_-\cdot P_k\right]\) by using the result from Eq.~\eqref{eq: final_fomula}. We can calculate it as follows:
\begin{align}
    &\mathrm{E}\left[ O_+ \cdot \rho_-\cdot P_k \cdot O_+ \cdot \rho_-\cdot P_k\right]\\=&\frac{1}{nd}\sum_{k=1}^{m}\mathrm{E}_{\theta_-, \theta_+}\left[ O_+ \cdot \rho_-\cdot P_k \cdot O_+ \cdot \rho_-\cdot P_k \right]\\
    =&\frac{1}{nd}\sum_{k=1}^{m}\mathrm{E}_{\theta_-}\left[ \mathrm{E}_{\theta_+}\left[ O_+ \cdot \rho_-\cdot P_k \cdot O_+\right] \cdot \rho_-\cdot P_k \right]\\
    =&\frac{1}{nd}\sum_{k=1}^{m}\mathrm{E}_{\theta_-}\left[ \frac{1}{4^n}\sum_{\sigma \in P(S)}\left\{\left(4^{|\sigma|-n}\right)^{d_+-1}\cdot I_\sigma^{\otimes|\sigma|}\otimes\text{Tr}_\sigma \left\{O^2\right\}\cdot \rho_- \cdot P_k\right\} \cdot \rho_-\cdot P_k \right]\\
    =&\frac{1}{4^nnd}\sum_{k=1}^{m}\sum_{\sigma \in \mathcal{P}(n)}\left\{\left(4^{|\sigma|-n}\right)^{d_+-1}\cdot I_\sigma^{\otimes|\sigma|}\otimes\text{Tr}_\sigma \left\{O^2\right\}\right\}\cdot \mathrm{E}_{\theta_-}\left[\rho_- \cdot P_k \cdot \rho_-\cdot P_k \right]\\
    =&\frac{1}{4^{2n}nd}\sum_{k=1}^{m}\sum_{\sigma,\tau \in \mathcal{P}(n)}\left\{\left(4^{|\sigma|-n}\right)^{d_+-1}\cdot I_\sigma^{\otimes|\sigma|}\otimes\text{Tr}_\sigma \left\{O^2\right\}\right\}\cdot \left\{\left(4^{|\tau|-n}\right)^{d_--1}\cdot I_\tau^{\otimes|\tau|}\otimes\text{Tr}_\tau \left\{\rho^2\right\}\cdot P_k^2\right\},
\end{align}
which \(d_-\) and \(d_+\) represents the depths of \(U_-\) and \(U_+\) respectively.

In deep parameterized quantum circuits, where \( d \) is sufficiently large, all coefficients of the form \(\left(4^{|\sigma|-n}\right)^{d_+-1}\) and \(\left(4^{|\tau|-n}\right)^{d_--1}\) will approach zero, except in the case where \( \sigma=\tau=\{1,2,\dots, n\} \).
So the expectation will be:
\begin{align}
    &\mathrm{E}\left[ O_+ \cdot \rho_-\cdot P_k \cdot O_+ \cdot \rho_-\cdot P_k\right]\\=&\frac{1}{4^{2n}nd}\sum_{k=1}^{m}\left\{I\cdot \text{Tr}\left\{O^2\right\}\text{Tr} \left\{\rho^2\right\}\right\}\\=&\frac{m}{4^{2n}nd}\text{Tr}\left\{O^2\right\}\text{Tr} \left\{\rho^2\right\}\cdot I.
\end{align}

\begin{figure}[h]
    \centering
    \begin{minipage}{0.48\textwidth}
        \centering
        \includegraphics[width=\textwidth]{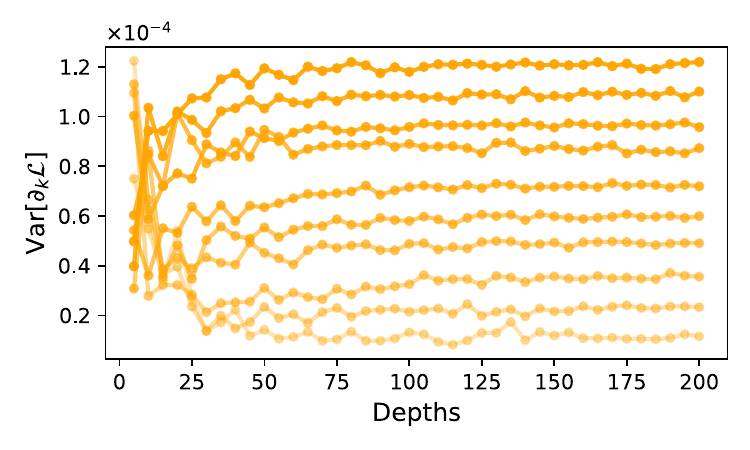}
    \end{minipage} \hfill
    \begin{minipage}{0.48\textwidth}
        \centering
        \includegraphics[width=\textwidth]{ratio.pdf}
    \end{minipage}
    \caption{This result is based on a 12-qubit PQCs, with the observable of \( Z^{\otimes 12} \). (a) shows the gradient variance from 5 to 200 depths, where the lines range from dark to light, representing the ratio between the effective number of parameters and the circuit depth from large to small. (b) shows the relationship between the gradient variance and the ratio between the effective number of parameters and the circuit depth at 200 layers.}
    \label{fig: supp}
\end{figure}

In summary, for deep parameterized quantum circuits, the gradient variance can be expressed as:
\begin{align}
    \mathrm{Var}[\partial_{k}\mathcal{L}]
    =&\frac{\alpha-1}{2}\text{Tr} \left\{\mathrm{E}\left[ O_+ \cdot \rho_-\cdot P \cdot O_+ \cdot \rho_-\cdot P\right]\right\}\\
    =&\frac{(\alpha-1)m}{2^{4n+1}nd}\text{Tr}\left\{ \text{Tr}\left\{O^2\right\}\text{Tr} \left\{\rho^2\right\}\cdot I \right\}\\
    =&\frac{(\alpha-1)m}{2^{3n+1}nd}\text{Tr}\left\{O^2\right\}\text{Tr} \left\{\rho^2\right\}\\
    \propto& \frac{m}{8^{n}nd},
\end{align}
where we use the fact that \(\Tr{I}=2^n\) and \(\Tr{O^2}=\Tr{\rho^2}=1\). When the number of depths is sufficiently deep, the variance value no longer changes, and the final result depends on the ratio between the effective number of parameters and the circuit depth, as shown in Fig.~\ref{fig: supp}. 

\bibliography{main}